\documentclass[12pt]{article}
\usepackage[utf8]{inputenc}
\usepackage[margin=1.0in]{geometry}
\usepackage{amssymb, amsmath, environ, tabularx,lipsum}
\usepackage{amsthm}
\usepackage{colordvi,verbatim,hyperref}
\usepackage{color,enumerate}
\usepackage{graphicx}
\usepackage{caption}
\usepackage{algorithm,algorithmic}
\usepackage{subcaption}
\usepackage[justification=centering]{caption}
\usepackage{authblk}
\usepackage[toc,page]{appendix}
\usepackage{lineno}

\makeatletter
\newcommand{\problemtitle}[1]{\gdef\@problemtitle{#1}}
\newcommand{\probleminput}[1]{\gdef\@probleminput{#1}}
\newcommand{\problemquestion}[1]{\gdef\@problemquestion{#1}}
\NewEnviron{problem}{
  \problemtitle{}\probleminput{}\problemquestion{}
  \BODY
  \par\addvspace{.5\baselineskip}
  \noindent
  \begin{tabularx}{\textwidth}{@{\hspace{\parindent}} l X c}
    \multicolumn{2}{@{\hspace{\parindent}}l}{\@problemtitle} \\
    \textbf{Input:} & \@probleminput \\
    \textbf{Output:} & \@problemquestion
  \end{tabularx}
  \par\addvspace{.5\baselineskip}
}
\makeatother

\theoremstyle{definition}

\newtheorem{lm}{Lemma}
\newtheorem{thrm}{Theorem}

\allowdisplaybreaks
\begin{document}

\title{Optimizing Edge Sets in Networks to Produce Ground Truth Communities Based on Modularity\footnotetext[0]{Email addresses: kosmad@rpi.edu (Daniel Kosmas), mitchj@rpi.edu (John E. Mitchell), tcshark@clemson.edu (Thomas C. Sharkey), szymab@rpi.edu (Boleslaw K. Szymanski)}}
\author[1]{Daniel Kosmas}
\author[1]{John E. Mitchell}
\author[2]{Thomas C. Sharkey} 
\author[3,4]{Boleslaw K. Szymanski}
\affil[1]{Department of Mathematical Sciences, Rensselaer Polytechnic Institute, Troy, NY 12180, USA}
\affil[2]{Department of Industrial Engineering, Clemson University, Clemson, SC 29634, USA}
\affil[3]{Department of Computer Science, Rensselaer Polytechnic Institute, Troy, NY 12180, USA}
\affil[4]{Spo\l{}eczna Akademia Nauk, \L{}\'{o}d\'{z}, Poland}
\date{}

\maketitle

\begin{abstract}
    We consider two new problems regarding the impact of edge addition or removal on the modularity of partitions (or community structures) in a network. The first problem seeks to add edges to enforce that a desired partition is the partition that maximizes modularity. The second problem seeks to find the sparsest representation of a network that has the same partition with maximum modularity as the original network. We present integer programming formulations, a row generation algorithm, and heuristic algorithms to solve these problems.  Further, we demonstrate a counter-intuitive behavior of modularity that makes the development of heuristics for general networks difficult. We then present results on a selection of social and illicit networks from the literature. \newline \footnotesize Keywords: community detection; modularity maximization; integer programming; network optimization; social networks; illicit networks
\end{abstract}

\section{Introduction}
Law enforcement agencies have been turning to social network analysis in order to augment their investigations \cite{sparrow1991application, van2009introduction, strang2014network, mohanty2020computational, cinar2017analyzing, anzoom2021review, cunningham2016understanding}. In particular, the problem of edge augmentation is of high importance to law enforcement; it is known that participants in illicit networks attempt to hide their activities from law enforcement in order to avoid detection \cite{berlusconi2013all, gill2013dynamic, spapens2011interaction}. However, missing edges can dramatically change the output of these quantitative tools, making the problem of identifying missing edges fundamentally important for the appropriate use of social network analysis in law enforcement investigations \cite{van2009introduction}. The problem of identifying missing edges in illicit networks has been approached from many different perspectives \cite{berlusconi2016link, calderoni2020robust, linkpred, rhodes2009inferring}. 

Another facet of using social network analysis in augmenting the investigations of illicit networks is through analyzing participant attributes. It has been identified that attributes of participants in illicit networks can provide direction for law enforcement investigations to understand the structure of these networks \cite{bright2013dismantling, malm2011networks, natarajan2000understanding, natarajan2006understanding}. In social network analysis, one way of uncovering potential node attributes is through community detection. The community detection problem seeks to partition the nodes into clusters, or communities, based on some measure \cite{modSurvey}. Calderoni et al.\ \cite{calderoni2017communities} identified that attributes, such as the roles of participants, are linked to the communities identified through modularity maximization \cite{mod_orig}. This connection has sparked further research into the application of modularity maximization to understanding illicit networks, based around the notion that incomplete information about networks impacts the communities identified \cite{Bahulkar2018, elsisy2020synthetic}. 

We introduce a new problem bridging the areas of edge augmentation and community detection. As law enforcement investigates an illicit network, they may not identify all of the edges \cite{spapens2011interaction, berlusconi2013all}. This, in turn, hinders the performance of community detection methods \cite{Bahulkar2018}, potentially causing the communities identified to disagree with the roles determined by the investigations. We seek to identify the minimum number of edges to add to the network to enforce that the roles determined by law enforcement investigations align with the communities determined via community detection. From here on, we refer to known communities from external investigation as the \emph{ground truth community structure}, or \emph{ground truth partition}.

We also consider a similar problem when law enforcement has observed parts of the illicit network, but is not yet confident about the roles of the participants. In this case, social network analysis is used to predict the roles of participants in an illicit network \cite{calderoni2017communities}, and law enforcement must then direct its investigations to validate these predictions. However, law enforcement is limited in its resources to perform all its duties \cite{kennedy2011risk, deeb2019understanding, holmes2008minority}. These resource constraints impact the quality of investigations \cite{cuganesan2011developments, goldstein2020exploitative}. In evaluating the quality of the communities determined via social network analysis, it is important that law enforcement validates these findings as efficiently as possible. This problem can be framed as identifying a sparse sub-network which maintains the partition of the original network. In other words, we seek to find the smallest number of edges (amongst the current ones) such that the partition that maximizes the modularity of the sub-network is the same as the partition that maximizes modularity of the original network. Understanding the minimum number of edges needed to produce these communities will allow scarce resources to be allocated to best investigate the edges of the network to prove or disprove what is believed by the investigations.

We present approaches to formulate these problems as mixed integer nonlinear programs (MINLPs).  For the case where edges are unweighted, we are able to formulate these problems as binary nonlinear programs. We then show how to reformulate the binary nonlinear program as an equivalent binary linear program.  For the case where edges are weighted, we show how the MINLP can be relaxed to a mixed integer linear program (MILP) and also discuss exact approaches to solve this problem. We additionally develop heuristic approaches to solve these problems, and demonstrate properties about modularity that make the development of more accurate heuristics difficult.  We apply our approaches to several case studies as a proof of concept and show that our heuristics perform quite well.

\subsection{Literature Review}
Community detection has been a problem of interest in network analysis for many years, with many different methods \cite{modSurvey}. Formally, consider a network $G = (V, E)$. A feasible partition of $V$ is defined to be a set of clusters, or communities, $P = (C_1, \ldots, C_k)$ such that $\bigcup_{i=1}^k C_i = V$ and $C_i \cap C_j = \emptyset$ for all $i \ne j$. Let $\mathcal{P}$ be the set of feasible partitions. The community detection problem seeks to find a partition $P^* \in \mathcal{P}$ that maximizes some measure of the partition. Typically, this measure would give preference to partitions where there are a large number of edges within each cluster, but a small number of edges between the clusters. Outside of illicit network analysis, community detection has applications in many domains, including the analysis of biological networks and social networks \cite{modSurvey}. 

A popular method to perform community detection is through modularity maximization \cite{chen2014community}, introduced by Newman and Girvan \cite{mod_orig}. Modularity maximization seeks to find a partition $P$ that maximizes the Newman-Girvan modularity metric, which compares, for each cluster, its number of actual edges and expected numbers of edges in the equivalent random network without communities \cite{newman2003random}. Modularity was initially defined on unweighted networks, but the definition can easily be extended to weighted networks. Brandes et al.\ \cite{brandes2007modularity} provides an equivalent definition of the unweighted version of modularity as 

\begin{equation*}
    Q = \sum_{C \in P}  \left(\frac{|E(C)|}{|E|}- \left(\frac{|E(C)| + \sum_{C' \in P} |E(C,C')|}{2|E|}\right)^2 \right)\text{,}
\end{equation*}
where $E(C)$ is the set of edges within cluster $C$ and $E(C,C')$ is the set of edges between clusters $C$ and $C'$. The problem of modularity maximization is known to be NP-hard \cite{brandes2007modularity}. Many heuristic methods have been developed to quickly find partitions with relatively large modularity scores \cite{newman2004fast, blondel2008fast}. Similar metrics have also been proposed to capture the same intuitive principles guiding modularity, while accounting for some of its shortcomings \cite{chen2013measuring}.

Modularity maximization can be formulated as an integer programming (IP) problem. Brandes et al.\ \cite{brandes2007modularity} introduces this formulation, and Agarwal and Kempe \cite{mod_IP} propose a rounding technique to improve the solve time of these problems. Given a weighted network $G = (V,E)$ with edge weights $w: E \rightarrow \mathbb{R}$, let $A$ be the weighted adjacency matrix of $G$, $m = \sum_{(i,j) \in E} w_{ij}$, and $d_i = \sum_{j \in V} w_{ij}$ for $i \in V$. For every pair of nodes $i,j \in V$, define decision variables $x_{ij}$ such that $x_{ij}=0$ if $i$ and $j$ are in the same cluster, and $x_{ij}=1$ otherwise. The integer programming formulation of modularity maximization is as follows:

\begin{align}
\label{mmip}
    \max_{x \in \{0,1\}^{|V|^2}} ~~~ & \frac{1}{2m}\sum_{i,j \in V} \left(A_{ij} - \frac{d_i d_j}{2m}\right)(1 - x_{ij}) \nonumber\\  
    \text{s.t.} \hspace{.050cm }& x_{ij} \le x_{ik} + x_{kj} & \text{ for all } i, j, k \in V
\end{align}

There have since been other formulations of this problem to improve solve time of the integer program (see \cite{aloise2010column} for such methods), but discussing them is outside the scope of our paper. Such methods can easily be adapted to fit into our framework.

One issue that arises in our problems, compared to this past work on using IP to maximize modularity, is the fact that we are enforcing that the modularity of the given partition is maximum amongst all partitions.  This leads to a number of constraints equal to the number of partitions, i.e., an exponential number of constraints based on the number of nodes in the network.  To avoid this difficulty, our framework operates on `row generation,' i.e., only maintaining a subset of these constraints at a time and then generating new constraints (rows) when needed.  In each iteration of our framework, a master problem is solved for a potential optimal solution based on a subset of the partition constraints. A subproblem is then solved to verify optimality, which fails when there is a partition not currently included that has a modularity larger than the ground truth community structure. In such a case, a new constraint is added to the master problem. Fischetti et al.\ \cite{monotone} demonstrates this is an effective method to solve bilevel integer optimization problems. More recently, community detection through the degree-corrected stochastic block model, which is known to be equivalent to modularity maximization \cite{newman2016equivalence}, has been modeled as a MINLP in \cite{serrano2021community}. The authors linearized the model and solved the resulting MILP using a row generation framework. However, our problem has the added complication of needing to compare the ground truth to all potential partitions in a network where the problem is deciding the existence of edges.  

\subsection{Paper Organization}
The rest of the paper is organized as follows: in Section 2, our two problems of interest are formally introduced. In Section 3, we describe a row generation algorithm to solve the problem exactly, and reformulate the problem as a MILP. In Section 4, we demonstrate properties of modularity that prove challenging when designing heuristic solution methods. We also present our heuristic approaches. In Section 5, we present computational results of our row generation algorithm and heuristics. We discuss our conclusions in Section 6.

\section{Description of Problems}
\subsection{Edge Addition for Community Optimization}
We first introduce the edge weight addition for community optimization problem. Consider a weighted network $G = (V,E)$ with weights $w: E \rightarrow \mathbb{R}$. Note that pairs of vertices that do not have an edge between them can be considered as having edges with weight $0$. Let $\mathcal{P}$ be the set of feasible partitions of $V$, with ground truth partition $T = (C_1^T, \ldots, C_k^T) \in \mathcal{P}$. Let $\bar{E}$ be the set of edges that can have weight added to the edge between them, and let $B_{ij}$ be the upper bound on the amount of weight that can be on edge $(i,j)$. Let $b_{ij}$ be the decision variable indicating how much weight is added to edge $(i,j)$. This problem considers adding weights to $G$ to enforce that the ground truth community structure $T = (C_1^T, \ldots,  C_k^T)$ is a partition that maximizes the modularity of the new network and $w_{ij} + b_{ij} \le B_{ij}$. 

\begin{problem}
  \label{prob:edgeAdd}
  \problemtitle{\textbf{Edge Weight Addition for Community Optimization}}
  \probleminput{A network $G = (V,E)$ with weights $w: E \rightarrow \mathbb{R}$, set of edges that can have weight added to them $\bar{E}$, weight limits on edges $B: E \cup \bar{E} \rightarrow \mathbb{R}_{\ge 0}$ and a specified partition $T = \{C_1^T, \ldots, C_k^T\}$}
  \problemquestion{Set of weights $b$ added to edges that enforce that $T$ is a partition that maximizes the modularity of the network $G$ with weights $w + b$.}
\end{problem}

Let $Q_P(w)$ be the modularity of $G$ with edge weights $w$. We can model the edge weight addition for community optimization problem as the following:

\begin{align}
    \label{model0w}
    \min_{b \in \mathbb{Z}_{\ge 0}}  ~~~ & \sum_{(i,j) \in \bar{E}} b_{ij} \nonumber\\  
    \text{s.t.} \hspace{.050cm }& T \in \text{arg}\max_{P \in \mathcal{P}} \{Q_P(w+b)\} & \\
    & w_{ij}+b_{ij} \le B_{ij} & \text{ for all } (i,j) \in \bar{E} \nonumber
\end{align}

By enumerating over all possible partitions, we can easily convert the constraint $T \in \text{arg}\max_{P \in \mathcal{P}} \{Q_P(b)\}$ in model \eqref{model0w} to the following equivalent model:

\begin{align}
    \label{model1w}
    \min_{b \in \mathbb{Z}_{\ge 0}}  ~~~ & \sum_{(i,j) \in \bar{E}} b_{ij} \nonumber\\  
    \text{s.t.} \hspace{.050cm }& Q_T(w+b) \ge Q_P(w+b) & \text{ for all } P \in \mathcal{P} \\
    & w_{ij}+b_{ij} \le B_{ij} & \text{ for all } (i,j) \in \bar{E} \nonumber
\end{align}

In this program, the constraint $Q_T(w+b) \ge Q_P(w+b)$ enforces that the modularity associated with partition $T$ after adding weights $b$ is at least as large as the modularity of every other partition $P$. Thus, $T$ is in the set of optimal partitions for the modularity maximization problem. To enforce that $T$ is the unique optimal solution, the constraint could be replaced with $Q_T(w+b) \ge Q_\mathcal{P}(w+b) + \epsilon$ for some small $\epsilon > 0$.

In the case where $G$ is unweighted, i.e., $w_{ij} = 1$ for all $(i,j) \in E$ and $w_{ij} = 0$ for all $(i,j) \notin E$, we can derive an equivalent model for model \eqref{model1w} that is an integer linear program. In this case, $\bar{E}$ will be the set of edges that can be added to the network, and $Q_P(E')$ will be the modularity of the partition $P$ in the network $G' = (V, E \cup E')$, where $E' = \{(i,j) \in \bar{E}: b_{ij}=1\}$. This special case can be modeled as:

\begin{align}
    \label{model1}
    \min_{E' \subseteq |\bar{E}|} ~~~ & |E'| \nonumber\\  
    \text{s.t.} \hspace{.050cm }& Q_T(E') \ge Q_P(E') & \text{ for all } P \in \mathcal{P} 
\end{align}

Since we can derive an integer linear program equivalent to model \eqref{model1}, which can be solved by commercial solvers, we will use this model as the basis for derivations in Section \ref{sec:ip}.

\subsection{Edge Removal for Community Preservation}
We now introduce the edge removal for community preservation problem. Consider a network $G = (V, E)$, and let $P^*$ be a partition that maximizes modularity of $G$. The edge removal for community preservation problem seeks to identify the smallest set $E' \subseteq E$ such that $P^*$ is a partition that maximizes modularity of $G' = (V,E')$.

\begin{problem}
  \label{prob:edgeRemove}
  \problemtitle{\textbf{Edge Removal for Community Preservation}}
  \probleminput{A network $G = (V,E)$, where $P^*$ is a partition that maximizes the modularity of $G$}
  \problemquestion{A network $G' = (V, E')$, where $E' \subset E$ and $P^*$ is a partition that maximizes the modularity of $G'$}
\end{problem}

This problem can be modeled similarly to model \eqref{model1}. We can express this as the following:

\begin{align}
    \label{model2}
    \min_{E' \subseteq E}  ~~~ & |E'|  \nonumber\\  
    \text{s.t.} \hspace{.050cm }& Q_{P^*} (E') \ge Q_P(E') & \text{ for all } P \in \mathcal{P}
\end{align}

Unlike model \eqref{model1}, where we optimize over an extended set of edges $\bar{E}$, here we optimize over the original set of edges $E$. Since we are trying to maintain that the original optimal partition $P^*$ is a partition that maximizes modularity, we use $P^*$ as the ground truth partition $T$ in the constraints. Because the constraints in both models are the same, we can use a similar solution method to solve both problems.

\section{Integer Programming Framework}
\label{sec:ip}
\subsection{Initial Framework}
To solve model \eqref{model1} by enumeration, we would need to verify the modularity of every possible partition, the number of which is exponentially large in the number of nodes. However, we expect that most of those partitions can be rejected based on modularity properties. For example, if every node is in the same cluster, or every node is in their own cluster, the modularity will always be zero. As long as our given partition has a non-negative modularity, we do not need to compare against these two partitions. 

We propose a row generation method to solve these problems. In this framework, we iteratively identify partitions of higher modularity than our given modularity. We then use these partitions to determine what edges should be added to the network so the given partition has a higher modularity than these partitions.

Let $\mathcal{P}^k$ be the set of partitions identified by the $k^{th}$ iteration of our iterative procedure. Instead of solving model \eqref{model1}, we solve the following:

\begin{align}
    \label{model1Reduced}
    \min_{z \in \{0,1\}^{|\bar{E}|}} ~~~ & \sum_{(i,j) \in \bar{E}} z_{ij} \nonumber\\  
    \text{s.t.} \hspace{.050cm }& Q_T(z) \ge Q_\mathcal{P}(z) & \text{ for all } P \in \mathcal{P}^k
\end{align}

In model \eqref{model1Reduced}, we enforce that $T$ has a modularity at least as large as every partition in $\mathcal{P}^k$, providing a lower bound on the objective value of model \eqref{model1}. However, this does not enforce that the given partition $T$ is in the set of partitions that maximize modularity. Let $z^k$ be the optimal solution found in model \eqref{model1Reduced}, and let $E^k$ be the set of edges added to the network with solution $z^k$, i.e., $E^k = \{(i,j) \in \bar{E}: z^k_{ij} = 1\}$. We now find a partition that maximizes modularity of the new network with edges added. Let $\bar{A}$ be the adjacency matrix of $G^k = (V, E \cup E^k)$, $\bar{m} = |E \cup E^k|$, and $\bar{d}_i$ be the degree of node $i$ in $G^k$. We find a partition that maximizes modularity with the following:

\begin{align}
\label{sp1}
    \max_{x \in \{0,1\}^{|V|^2}} ~~~ & \frac{1}{2\bar{m}}\sum_{i,j \in V} \left(\bar{A}_{ij} - \frac{\bar{d}_i\bar{d}_j}{2\bar{m}}\right)(1 - x_{ij}) \nonumber\\  
    \text{s.t.} \hspace{.050cm }& x_{ij} \le x_{ik} + x_{kj} & \text{ for all } i, j, k \in V
\end{align}

By solving model \eqref{sp1}, we find a partition $\bar{P}$ that maximizes modularity of the new network, as well as its modularity. If $Q_T (z^k) = Q_{\bar{P}} (z^k)$, then we have that $T$ is in the set of optimal partitions, and we are done. If not, then we have found a partition, $\bar{P}$, which has greater modularity than $T$ which must be considered when adding edges. We set $\mathcal{P}^{k+1} = \mathcal{P}^k \cup \bar{P}$, and repeat this process. Formalizing this procedure yields Algorithm \ref{alg:alg1}.

\begin{algorithm}
\caption{Row Generation for Edge Addition}
\label{alg:alg1}
\begin{algorithmic}
\STATE{\textbf{Initialize:}} network $G = (V, E)$, ground truth partition $T$, initial set of partitions $\mathcal{P}^1$, set of includable edges $\bar{E}$, iteration counter $k=1$.
 \WHILE{not converged}
\STATE{\textbf{Step 1.}} Solve model \eqref{model1Reduced} for $z^k$, $E^k$.
\IF{\eqref{model1Reduced} is infeasible}
 \STATE{terminate; model \eqref{model1} is infeasible.}
\ENDIF
\STATE{\textbf{Step 2.}} Create test network $G^k = (V, E \cup E^k)$. With test network $G^k$, solve model \eqref{sp1} for optimal partition $\bar{P}$.
\IF{$Q_{T} (z^k) = Q_{\bar{P}} (z^k)$}
 \STATE{terminate; $E^k$ is the optimal solution to model \eqref{model1}}.
\ELSE
 \STATE{Set $\mathcal{P}^{k+1} = \mathcal{P}^k \cup \bar{P}$, $k \leftarrow k+1$, and return to Step 1.}
\ENDIF
\ENDWHILE
\end{algorithmic}
\end{algorithm}

To analyze, each iteration of this procedure can result in one of three outcomes. First, if model \eqref{model1Reduced} is infeasible, then model \eqref{model1} is also infeasible. If model \eqref{model1Reduced} is feasible, we have two cases. If $Q_T (z^k) = Q_{\bar{P}} (z^k)$ in the $k^{th}$ iteration, we have found an optimal set of edges to add to the network.  Otherwise, we have $Q_T (z^k) < Q_{\bar{P}} (z^k)$, so  we include $\bar{P}$ in $\mathcal{P}^{k+1}$, and we continue the process. Since $|\mathcal{P}|$ is finite, our framework has a finite number of iterations, and will thus solve model \eqref{model1} in finite time.

We note that modularity maximization is known to be NP-Hard, meaning solving model \eqref{sp1} in Algorithm \ref{alg:alg1} involves solving an NP-Hard problem each iteration. It may, however, not be necessary to solve model \eqref{sp1} to optimality; the purpose of solving model \eqref{sp1} is to determine if the partition $T$ is optimal in $G^k$. If it is not optimal, we wish to find a partition $\bar{P}$ with $Q_{\bar{P}}(z^k) > Q_{T}(z^k)$, not necessarily the partition that maximizes modularity of $G^k$. As long as a partition with $\bar{P}$ with $Q_{\bar{P}}(z^k) > Q_{T}(z^k)$ is identified, Algorithm \ref{alg:alg1} will still converge to the true solution. Alternatively, solving model \eqref{sp1} may be replaced with other methods, such as a heuristic method for identifying partitions with near-optimal modularity. In doing so, Algorithm \ref{alg:alg1} provides a lower bound on the objective value, since not solving model \eqref{sp1} to optimality may result in claiming that no partition exists such that $Q_{P}(z^k) \ge Q_T(z^k)$ even when such a partition might exist. However, for the networks we test our methods on, it was found that solving model \eqref{model1Reduced} was more computationally challenging than solving model \eqref{sp1}. We elaborate further on computational testing in Section 5.

While we choose to use modularity as our measure of community structure, note that our framework can work for any optimization-based clustering measure. In model \eqref{model1}, and likewise model \eqref{model1Reduced}, the constraint enforcing that the given partition $T$ has modularity at least as large as that of all other partitions would be replaced with the desired measure of $T$ being at least as large as that of all other partitions, and model \eqref{sp1} would be replaced with finding a partition that optimizes that measure.  In this paper, though, we are able to express model \eqref{model1} as an integer program, rather than a nonlinear integer program, which we demonstrate over the next two subsections.

\subsection{Nonlinear Integer Programming Representation}
We now express model \eqref{model1} as an integer program. We can rewrite the modularity of a given partition with added edges as:
\begin{equation}
    Q_P(z) = \sum_{i,j \in V} \left(\frac{A_{ij} + z_{ij}}{2(m+\sum_{(s,t)\in \bar{E}}z_{st})} - \frac{(d_i + \sum_{l\in V}z_{il})(d_j + \sum_{h \in V}z_{jh})}{(2(m+\sum_{(s,t)\in \bar{E}}z_{st}))^2}\right)(1-x^P_{ij})\text{.}
\end{equation}

We can thus express the nonlinear integer program as:
\small\begin{align}
    \label{model1IP}
    \min_{z \in \{0, 1\}^{|\bar{E}}|}  & \sum_{(i,j) \in \bar{E}} z_{ij} \nonumber\\  
    \text{s.t.} \hspace{.050cm }  & \sum_{i,j \in V} \left(\frac{A_{ij} + z_{ij}}{2(m+\sum_{(i,j)\in \bar{E}}z_{ij})} - \frac{(d_i + \sum_{l\in V}z_{il})(d_j + \sum_{h \in V}z_{jh})}{(2(m+\sum_{(s,t)\in \bar{E}}z_{st}))^2}\right)(1-x^T_{ij}) \\
    &\ge \sum_{i,j \in V} \left(\frac{A_{ij} + z_{ij}}{2(m+\sum_{(i,j)\in \bar{E}}z_{ij})} - \frac{(d_i + \sum_{l\in V}z_{il})(d_j + \sum_{h \in V}z_{jh})}{(2(m+\sum_{(s,t)\in \bar{E}}z_{st}))^2}\right)(1-x^P_{ij}) \nonumber \\ & \text{ for all } P \in \mathcal{P} \nonumber 
\end{align}

We note that model \eqref{model2} can be expressed similarly, where $z_{ij}$ is no longer a binary vector indicating inclusion of edges, but a vector of integers, expressing the amount of weight added to edges.

\subsection{Linearizing the Problem}
The constraints in model \eqref{model1IP} are nonlinear, but, by using standard techniques, we can create linear constraints that are equivalent. Below, we describe the process of linearizing these constraints, which we summarize in Theorem 1. The equivalent constraint in the weighted version of the problem can be linearized similarly, but we will highlight some key differences after the derivation is complete.

The first point to note is that the $2(m + \sum_{(i,j) \in \bar{E}}z_{ij})$ in the denominator is the same on both sides of the equality. Furthermore, it will always be strictly greater than $0$, assuming the network has a nonempty edge set. Thus, multiplying both sides of the inequality by this denominator does not change the inequality, resulting in: 
\begin{align}
    \label{step1}
    & \sum_{i,j \in V} \left(A_{ij} + z_{ij} - \frac{(d_i + \sum_{l\in V}z_{il})(d_j + \sum_{h \in V}z_{jh})}{2(m+\sum_{(s,t)\in \bar{E}}z_{st})}\right)(1-x^T_{ij}) \\
    &\ge \sum_{i,j \in V} \left(A_{ij} + z_{ij} - \frac{(d_i + \sum_{l\in V}z_{il})(d_j + \sum_{h \in V}z_{jh})}{2(m+\sum_{(s,t)\in \bar{E}}z_{st})}\right)(1-x^P_{ij})\text{.} \nonumber 
\end{align}

Again, using that $2(m + \sum_{(i,j) \in \bar{E}}z_{ij}) > 0$, we multiply both sides of the inequality by $2(m + \sum_{(i,j) \in \bar{E}}z_{ij})$ to remove fractions from our inequality. Additionally, the factors $1 - x_{ij}^T$ and $1 - x_{ij}^P$ are multiplied by the same terms. Thus, we can equivalently express the inequality as the following:

\begin{align}
    \label{step2}
    &\sum_{i,j \in V}\left( (A_{ij} + z_{ij})(2(m+\sum_{(s,t)\in \bar{E}}z_{st})) - (d_i + \sum_{l\in V}z_{il})(d_j + \sum_{h \in V}z_{jh})\right)(x^P_{ij} - x^T_{ij}) \ge 0
\end{align}

We now have a quadratic constraint equivalent to the original constraint. To linearize this constraint, we use the McCormick inequalities \cite{mccormick1976computability}. For a pair of edges $(i,j)$ and $(s,t)$, let $w_{ijst} = z_{ij} z_{st}$. The following inequalities enforce that $w_{ijuv}$ takes the correct value:

\begin{align*}
    w_{ijst} &\le z_{ij}\text{,}\\
    w_{ijst} &\le z_{st}\text{,}\\
    w_{ijst} &\ge 0 \text{,}\\
    w_{ijst} &\ge z_{ij} + z_{st} - 1\text{.}
\end{align*}

By multiplying out the quadratic terms, and replacing any bilinear terms with the appropriate variable, we can express the term multiplied by $x_{ij}^T$ (and $x_{ij}^P$) as 

\begin{align*}
    f_{ij} (z, w) &= 2mA_{ij}+2A_{ij}\sum_{(s,t)\in \bar{E}} z_{st} + 2m z_{ij} + 2\sum_{(s,t) \in \bar{E}} w_{ijst} \\ 
    & - \left(d_i d_j + d_i\sum_{h\in V}z_{jh} + d_j\sum_{l \in V}z_{il} + \sum_{h,l \in V} w_{iljh}\right)\text{.}
\end{align*} 

We can now express the linearized constraint as:

\begin{align}
\label{constraint}
    \sum_{i,j \in V} (x_{ij}^P - x_{ij}^T) f_{ij}(z,w) \ge 0
\end{align}

Since the $z$ variables are binary, we will have that $w_{ijst} = z_{ij} z_{st}$. Thus, our integer programming formulation is:

\begin{align} 
    \label{model1IPFinal}
    \min_{z \in \{0, 1\}^{|\bar{E}|}, w}  ~~~ & \sum_{(i,j) \in \bar{E}} z_{ij} \nonumber\\  
    \text{s.t.} \hspace{.050cm }& \sum_{i,j \in V} (x_{ij}^P - x_{ij}^T) f_{ij}(z,w) \ge 0 \text{ for all } P \in \mathcal{P}\\
    & w_{ijst} \le z_{ij}\text{ for all } (i,j), (s,t) \nonumber\\
    & w_{ijst} \le z_{st}\text{ for all } (i,j), (s,t) \nonumber\\
    & w_{ijst} \ge 0 \text{ for all } (i,j), (s,t) \nonumber\\
    & w_{ijst} \ge z_{ij} + z_{st} - 1\text{ for all } (i,j), (s,t) \nonumber
\end{align}

\begin{thrm}
Model \eqref{model1IPFinal} is equivalent to model \eqref{model1IP}: there is a one-to-one correspondence between solutions for \eqref{model1IP} and model \eqref{model1IPFinal}.
\end{thrm}
\begin{proof}
We show that model \eqref{model1} is equivalent to model \eqref{model1IPFinal} by showing that a feasible solution to one model has a corresponding feasible solution to the other model with the same objective value. Note that, since the objective function value of both models is the same, as long as $z$ remains unchanged between the two models, the objective value of the solution will also be unchanged. 

Suppose $(z, w)$ is a feasible solution to model \eqref{model1IPFinal}. Since the constraints of model \eqref{model1IP} are equivalent to constraints \eqref{constraint}, and $(z,w)$ is a feasible solution to model \eqref{model1IPFinal}, $z$ is a feasible solution to model \eqref{model1IP} with equivalent objective value. Likewise, suppose $z$ is a feasible solution to model \eqref{model1IP}. For pairs of edges $(i,j), (s,t)$, compute $w_{ijst} = z_{ij} z_{st}$. Suppose, without loss of generality, $z_{ij} = 0$, and thus $w_{ijst} = 0$. The McCormick inequalities will be satisfied, as $0 = w_{ijst} = z_{ij} \le z_{st}$ and $w_{ijst} = 0 = z_{ij} \ge z_{ij} + z_{st} - 1$. Now suppose $z_{ij}=z_{st}=1$. The McCormick inequalities are again satisfied, since $w_{ijst} = 1 \le z_{ij} = z_{st}$ and $w_{ijst} = 1 \ge 1 = z_{ij} + z_{st} - 1$. Additionally, since the constraints of model \eqref{model1IP} are equivalent to constraints \eqref{constraint}, $(z,w)$ is a feasible solution to model \eqref{model1IPFinal} with equivalent objective value. Thus, model \eqref{model1IP} is equivalent to model \eqref{model1IPFinal}.
\end{proof}

In the framework of Algorithm \ref{alg:alg1}, we would solve model \eqref{model1IPFinal} to determine $E^k$, then solve model \eqref{sp1} to determine a partition that maximizes modularity of the network $G^k = (V, E\cup E^k)$. We additionally add a constraint regarding the previous iteration's objective value to model \eqref{model1IPFinal}. Let $c^k$ be the objective value in the $k^{th}$ iteration. We add the constraint $\sum_{(i,j) \in K_{|V|} \setminus E} z_{ij} \ge c^k$, with $c^0$ initialized as $0$. By enforcing that the objective value is non-decreasing, we can reduce the time of the branch-and-bound procedure needed to solve iterations where the objective value stays the same as in the previous iteration.

We note that, when extending these manipulations to the weighted version of the problem, we use the generalized McCormick envelope \cite{mccormick1976computability}. For integer variables $z_{ij}, z_{st}$ with $l_{ij} \le z_{ij} \le u_{ij}$ and $l_{st} \le z_{st} \le u_{st}$, we can represent $w_{ijst}$ as:

\begin{align*}
    w_{ijst} \le l_{st} z_{ij} + u_{ij} z_{st} - l_{st} u_{ij}\\
    w_{ijst} \le l_{ij} z_{st} + u_{st} z_{ij} - l_{ij} u_{st}\\
    w_{ijst} \ge l_{st} z_{ij} + l_{ij} z_{st} - l_{ij} l_{st}\\
    w_{ijst} \ge u_{st} z_{ij} + l_{ij} z_{st} - u_{ij} u_{st}
\end{align*}

However, for non-binary $z_{ij}$ and $z_{st}$, we no longer have the guarantee that $w_{ijst} = z_{ij}z_{st}$, and these substitutions will not result in an equivalent model. A potential solution to this is to use binary expansion, replacing each integer variable with a set of binary variables to represent any value that integer variable can take \cite{owen2002value}. More formally, for integer variable $z_{ij}$ with $z_{ij} \le u_{ij}$, we would create binary variables $b_{ij}^0, \ldots, b_{ij}^k$, with $k = \lfloor \log_2(u_{ij}) \rfloor$. We would then include $z_{ij} = \sum_{l=0}^k 2^l b_{ij}^l$, and replace $z_{ij}$ with $\sum_{l=0}^k 2^{l} b_{ij}^l$ in the model, converting the bilinear terms with integer variables to bilinear terms with binary variables. While this substitution would result in an equivalent model, it also results in a large number of binary variables that does not scale well with problem size. Preliminary computational results show this is not a feasible approach.

\subsection{Disjunctive Cuts and Additional Partitions}
\label{sec:disj}
In order to better enforce that the ground truth partition is optimal, we augment model \eqref{model1Reduced} by defining valid disjunctive cuts, and augment Algorithm \ref{alg:alg1} by generating additional partitions. We first consider how to enforce that a single node $v$ is clustered correctly. For ease of notation, consider a pair of arbitrary partitions $P = (C_1 \cup \{v\}, C_2, C_3, \ldots, C_l)$ and $P' = (C_1, C_2 \cup\{v\}, C_3, \ldots, C_l)$, where $P$ is the desired partition. Let $e_i (v)$ be the set of edges between $v$ and cluster $C_i$ for $i = 1, 2$ and let $e_{-} (v)$ be the set of edges between $v$ and $\bigcup_{j=3}^l C_j$. Let $E(C_i, C_{-})$ be the set of edges between $C_i$ and $\bigcup_{j=3}^l C_j$, and let $Q_{-}$ be the contribution to modularity by $C_3, \ldots, C_l$. We want to identify conditions on when $Q_P < Q_{P'}$ to derive valid inequalities that can be added to the model \eqref{model1IPFinal} to prevent it from occurring. 

\begin{align*}
    & (4m^2)\left(Q_P - Q_{P'}\right) \\
    & = 4m(|E(C_1)| + |e_1(v)|) - \left(2|E(C_1)| + 2 |e_1(v)| + |E (C_1, C_2)| + |e_2 (v)| + |E(C_1, C_{-})| + |e_{-}(v)|\right)^2\\
    & + 4m|E(C_2)| - \left(|2E(C_2)| + |E (C_1, C_2)| + |e_2 (v)| + |E(C_2, C_{-})|\right)^2 + 4m Q_{-}\\
    & - 4m|E(C_1)| + \left(2|E(C_1)| + |E (C_1, C_2)| + |e_1 (v)| + |E(C_1, C_{-})|\right)^2 - 4m Q_{-}\\
    & - 4m(|E(C_2)| + |e_2 (v)|) + \left(2E(C_2) + 2|e_2 (v)| + |E (C_1, C_2)| + |e_1 (v)| + |E(C_2, C_{-})| + |e_{-} (v)|\right)^2\\
    & = 4m(|e_1 (v)| - |e_2 (v)|) + \left(2|E(C_1)| + |e_1 (v)| + |E (C_1, C_2)| + |E(C_1, C_{-})|\right)^2 \\
    & - [(2|E(C_1)| + |e_1(v)| + |E (C_1, C_2)| + |E(C_1, C_{-})|) + (|e_1 (v)| + |e_2 (v)|  + |e_{-}(v)|)]^2 \\
    & - \left(|2E(C_2)| + |e_2 (v)| + |E (C_1, C_2)| + |E(C_2, C_{-})|\right)^2\\
    & + [(2|E(C_2)| + |e_2(v)| + |E (C_1, C_2)| + |E(C_2, C_{-})|) + (|e_1 (v)| + |e_2 (v)|  + |e_{-}(v)|)]^2\\
    & = 4m(|e_1 (v)| - |e_2 (v)|) - (|e_1 (v)| + |e_2 (v)| + |e_{-} (v)|)^2\\
    & - 2\left(2|E(C_1)| + |e_1 (v)| + |E(C_1,C_2)| + |E(C_1, C_{-})|\right)\left(|e_1 (v)| + |e_2 (v)| + |e_{-} (v)|\right) \\
    & + 2\left(2|E(C_2)| + |e_2 (v)| + |E(C_1,C_2)| + |E(C_2, C_{-})|\right)\left(|e_1 (v)| + |e_2 (v)| + |e_{-} (v)|\right) \\
    & + (|e_1 (v)| + |e_2 (v)| + |e_{-} (v)|)^2 \\
    & = (4m - 2|e_1 (v)| - 2|e_2 (v)| - 2|e_{-} (v)|)(|e_1 (v)| - |e_2 (v)|) \\
    & - 2(2|E(C_1)| + |E(C_1, C_{-})| - 2|E(C_2)| - |E(C_2, C_{-})|)(|e_1 (v)| + |e_2 (v)| + |e_{-} (v)|)
\end{align*}

Note that $0 \le |e_1 (v)| + |e_2 (v)| + |e_{-} (v)| \le m$, so if $|e_1 (v)| \le |e_2(v)|$ and $2|E(C_1)| + |E(C_1, C_{-})| > 2|E(C_2)| + |E(C_2, C_{-})|$, then $Q_P < Q_{P'}$. Thus, we must have that 

\begin{equation*}
    |e_1 (v)| \ge |e_2 (v)| + 1 \text{ or } 2|E(C_2)| + |(E(C_2,C_{-})| \ge 2|E(C_1)| + |(E(C_1,C_{-})|\text{.}
\end{equation*}

Likewise, if $|e_1 (v)| < |e_2(v)|$ and $2|E(C_1)| + |E(C_1, C_{-})| \ge 2|E(C_2)| + |E(C_2, C_{-})|$, then $Q_P < Q_{P'}$. This results in needing to enforce 

\begin{equation*}
    e_1 (v)| \ge |e_2 (v)| \text{ or } 2|E(C_2)| + |(E(C_2,C_{-})| \ge 2|E(C_1)| + |(E(C_1,C_{-})| + 1\text{.}
\end{equation*}

To define these cuts regarding $T$, we must first discuss how to determine which nodes are ``classified incorrectly," especially when our ground truth community structure and the current optimal partition have a different number of clusters. Let $\bar{P}$ be a partition that maximizes modularity, as determined by Algorithm \ref{alg:alg1}. We construct a complete bipartite network, where one set of nodes represents the clusters in $T$, and the other set represents the clusters in $\bar{P}$. For a pair of clusters $C_i^T$ and $C_j^{\bar{P}}$, we set the edge weight for $(i,j)$ to be $|C^T_i \cap C_j^{\bar{P}}|$, the number of nodes shared by those clusters. We then solve the maximum weight matching problem on this bipartite network, indicating how the clusters should be ordered. From this, we can easily determine which nodes are incorrectly classified. Note that if an optimal partition has more clusters than the ground truth community structure, then at least one cluster in an optimal partition will not be matched with any cluster in the ground truth. In this scenario, all nodes in this cluster are considered incorrectly classified. Similarly, if the ground truth community structure has more clusters than an optimal partition, at least one cluster in the ground truth communities will not be matched with any cluster in an optimal partition. All nodes in this unmatched cluster will also be considered incorrectly classified. An example of this distinction being important is when two clusters in the ground truth community structure are merged into a single cluster in an optimal partition. While these nodes will all be clustered with all of the nodes they should be clustered with, whichever ground truth cluster is smaller will have its nodes considered incorrectly classified.  This indicates that edges (or weights) must be added to distinguish this smaller cluster as its own individual cluster.

Once we have determined the set of incorrectly classified nodes $M$, we can now define the disjunctive cuts. Let $v \in M$ be an incorrectly classified node, and, without loss of generality, assume $v \in C_1^T$. Let $\bar{C}$ be the cluster $v$ belongs to in $\bar{P}$. Suppose $\bar{C} \setminus C_1^T$ is nonempty, and consider $w \in \bar{C} \setminus C_1^T$. Without loss of generality, assume $w \in C_2^T$. We define the partition $P' = (C_1^T \setminus \{v\}, C_2^T \cup \{v\}, C_3^T, \ldots, C_l^T)$. By utilizing $\bar{P}$, we ensure that the partition generated provides a meaningful contribution by clustering $v$ with a node with which it was previously clustered. Besides defining the disjunctive cuts, we also include the partition $P'$ in $\mathcal{P}^k$. Computationally, this inclusion improved solve time. 

In the case when $\bar{C} \setminus C_1^T = \emptyset$, the above partition is no longer properly defined, as $\bar{C} \subseteq C_1^T$. One option would be to have $v$ in a cluster on its own in $P'$. This partition, however, is known to not be optimal as long as $d_v > 0$ \cite{brandes2007modularity}. Instead, we consider a partition that separates all of $\bar{C}$, not just a single node, since $\bar{C} \subset C_1^T$. We can define a new partition $P' = (C_1^T \setminus \bar{C}, \bar{C}, C_2^T, \ldots, C_l^T)$. This partition is well defined, since if $C_1^T \setminus \bar{C} = \emptyset$ and $\bar{C} \setminus C_1^T = \emptyset$, then $C_1^T = \bar{C}$ and thus $v \notin M$. By including the constraint $Q_T \ge Q_{P'}$, we enforce that edges are added to merge $C_1^T \setminus \bar{C}$ and $\bar{C}$.

At each iteration of Algorithm \ref{alg:alg1}, we generate the set of incorrectly classified nodes $M$, and remove nodes whose cluster in $\bar{P}$ is a subset of a cluster in $T$. For each of these nodes $v$, we pick an arbitrary node $w$, where $v$ and $w$ are in the same cluster in $\bar{P}$, but different clusters in $T$, to define the disjunctive cuts and additional partition. Then, for every cluster $C_i^{\bar{P}} \in \bar{P}$, we check if it is a subset of a cluster in $C_j^T \in T$, and generate additional partitions for it when $C_i^{\bar{P}} \subset C_j^T$.

\section{Challenges for Heuristic Approaches}
The potential downsides of using model \eqref{model1IPFinal} include (i) that there could be a large number of binary variables, particularly if the network is sparse, and (ii) we must repeatedly solve model \eqref{model1IPFinal} as we generate new partitions to be added to it. With this, even small networks can be computationally expensive to solve, especially when trying to solve model \eqref{model2}. Thus, good heuristics are important for obtaining high quality solutions in a reasonable amount of time. Before we present our heuristic algorithms, we discuss some counter-intuitive behaviors of modularity as a measure of community strength.

\subsection{Counter-Intuitive Behavior of Modularity}
\begin{figure}[ht]
        \begin{minipage}[b]{0.49\linewidth}
            \centering
            \captionsetup{justification=centering}
            \includegraphics[width=\textwidth]{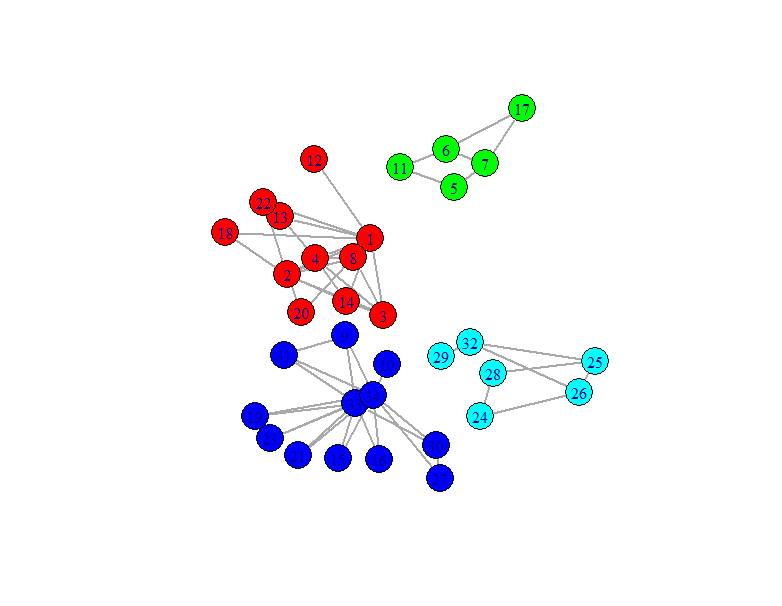}
            \caption*{$0.6753$ Modularity}
            \label{fig:a1}
        \end{minipage}
        \begin{minipage}[b]{0.49\linewidth}
            \centering
            \captionsetup{justification=centering}
            \includegraphics[width=\textwidth]{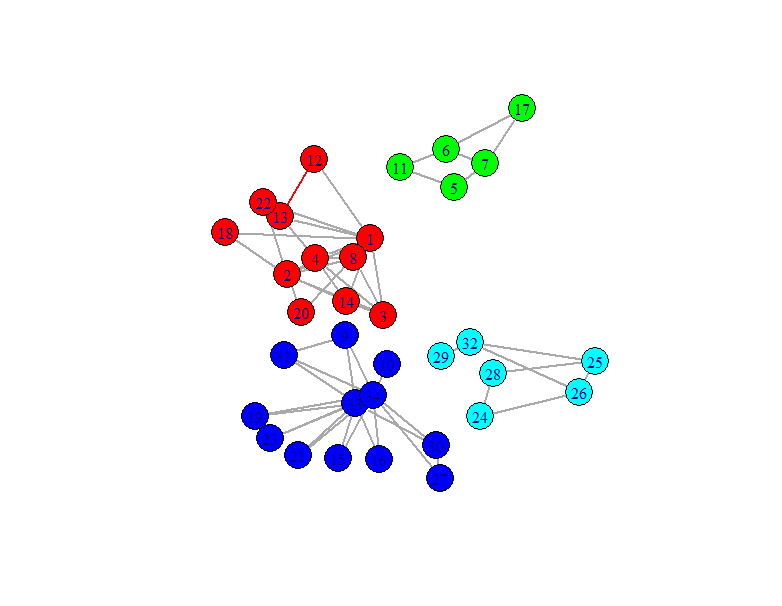}
            \caption*{Add red edge: $0.6724$ Modularity}
            \label{fig:b1}
        \end{minipage}
        \caption{Example of adding an edge within a cluster decreasing modularity of the partition}
        \label{decrease}
\end{figure}

As modularity is a measure of the strength of communities, there are two nice properties we would hope to have. Intuitively, we would expect that adding an edge between two nodes in the same cluster would increase the modularity score of that partition. Likewise, we would expect that adding an edge between two nodes in different clusters would decrease the modularity score of that partition. Neither of these intuitive properties are true. Figure \ref{decrease} demonstrates that adding an edge within a cluster can decrease modularity. 

We construct this example by taking Zachary's karate club network \cite{zachary} and removing all edges between distinct communities in an optimal partition of the network. The modularity of the partition where each disjoint component is its own community is $0.6753$. By adding the red edge, which is between two nodes within the red cluster, the modularity decreases to $0.6724$. 

\begin{figure}[ht]
        \begin{minipage}[b]{0.49\linewidth}
            \centering
            \captionsetup{justification=centering}
            \includegraphics[width=\textwidth]{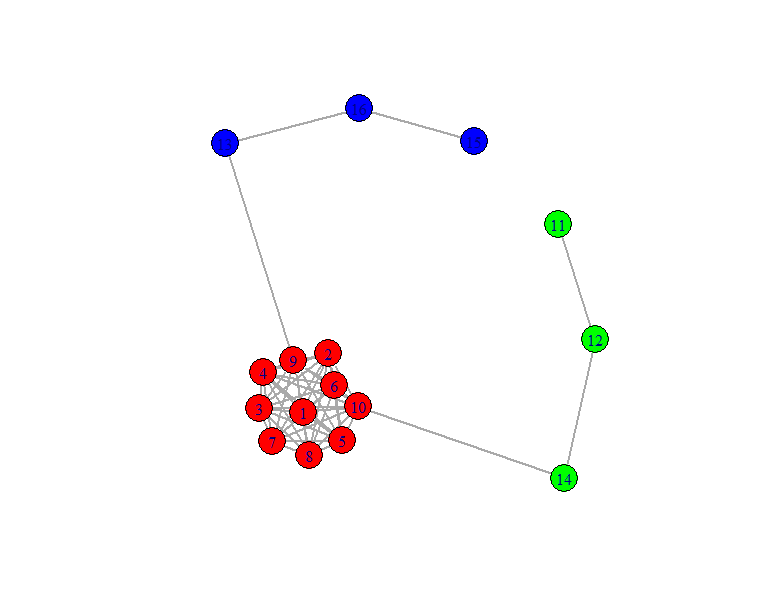}
            \caption*{$0.1424$ Modularity}
            \label{fig:a2}
        \end{minipage}
        \begin{minipage}[b]{0.49\linewidth}
            \centering
            \captionsetup{justification=centering}
            \includegraphics[width=\textwidth]{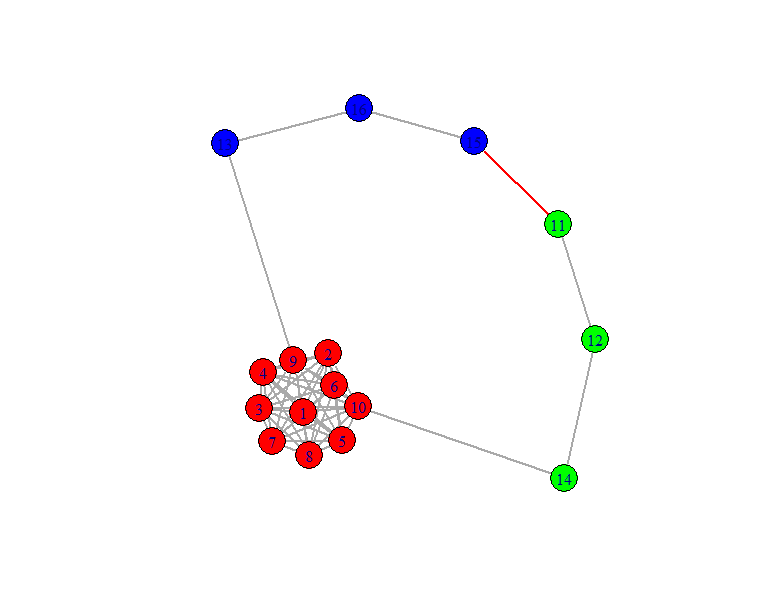}
            \caption*{Add red edge: $0.1531$ Modularity}
            \label{fig:b2}
        \end{minipage}
        \caption{Example of adding an edge across clusters increasing modularity of the partition}
        \label{increase}
\end{figure}

We can likewise create an example where adding an edge across clusters increases the modularity. Figure \ref{increase} demonstrates this behavior.

We create the example by starting with a complete network on $10$ nodes. We then add six more nodes, and create two disjoint paths of edge-length two on these nodes. We take one endpoint of each path, and connect them to one node each in the complete sub-network. With the partition where the complete sub-network and two paths are in their own clusters, we have a modularity of $0.1424$. By adding the red edge, which connects the degree $1$ nodes in the blue and green clusters, the modularity increases to $0.1531$.

A property of modularity being a global measure is that only the numbers of edges within clusters and between clusters are relevant to computing modularity, but not what specific nodes each edge connects \cite{mod_orig}. We would then expect that the change in modularity of a particular partition from adding an edge is dependent only on the clusters the incident nodes are in, not the nodes themselves. Exact computations for the change in modularity are presented in Appendix \ref{app:proofs}. 

While this is a nice property for a global measure, this property makes developing heuristics based on local properties difficult; we cannot distinguish between edges to add to improve the modularity of our ground truth communities. Further, the change in modularity \emph{across different partitions} will be different and thus understanding how adding an edge changes the current best partition remains challenging without enumerating all partitions. However, we still use our intuition in developing our heuristics.

We also wish to understand what edges may be necessary in creating a distinct cluster. From Brandes et al.\ \cite{brandes2007modularity}, we know the following two properties.
\begin{lm}
\label{lm:connect}
A partition $P$ that maximizes modularity will not have disconnected clusters.
\end{lm}

\begin{lm}
\label{lm:deg1}
A partition $P$ that maximizes modularity will not have any clusters that consist of a single node with degree $1$.
\end{lm}

We can combine these two results into a more powerful result regarding star networks \cite{pemmaraju2003computational}.

\begin{thrm}
\label{thrm:star}
Given a partition $P = \{C_1, \ldots, C_k\}$, a sparsest network, where each cluster is connected, to have $P$ as a partition that maximizes modularity has each cluster $C_i$ consisting of edges that form a star network.
\end{thrm}
\begin{proof}
Consider a network $G$ on $n$ nodes, and partition $P = \{C_1, \ldots, C_k\}$. In order to have cluster $i$ constitute a connected component, we need to have at least $|C_i| - 1$ edges. Thus, to have that every cluster is a connected component, we need to have at least $n-k$ edges. We show that we can do so by demonstrating that if each cluster is a star network, then $P$ is a partition that maximizes modularity.

Suppose the edges of $G$ induce a star network on each cluster. Note that, by Lemma \ref{lm:connect}, nodes from distinct star networks will not be clustered together in an optimal partition. Furthermore, also by Lemma \ref{lm:connect}, in a partition that maximizes modularity, multiple degree $1$ nodes in the $i^{th}$ star network will not be clustered together unless they are also clustered with the degree $|C_i - 1|$ node. Thus, every center node in a star network will be in a distinct cluster, and every degree $1$ node will be in its own cluster or with the center node of its star network. However, by \ref{lm:deg1}, if there is a degree $1$ node in its own cluster, that partition is not optimal. Thus, the only partition that maximizes the modularity of $G$ will have the $i^{th}$ cluster as the $i^{th}$ star network. Since $|E(G)| = n - k$, there exists a network with $n-k$ edges that has $P$ as a partition that maximizes modularity.
\end{proof}

This theorem helps us understand what edges would be useful to add to the network to get the best gain for the smallest number of edges. In particular, we will seek to `reinforce' star-like sub-networks for each distinct cluster in the ground truth. This also provides a lower bound on the objective value of model \eqref{model2}.

\subsection{Heuristic Algorithm for Edge Addition}
Since we know exactly how adding edges to the network will impact modularity, we can easily describe the ideal edges to add: ones that are within a single cluster in ground truth partition $T$, and between two clusters in the currently optimal partition $P^*$. Our procedure focuses on adding edges (or weights) to the network that increase the modularity of $T$. The procedure is:
\begin{enumerate}
    \item Identify the ``most incorrectly classified" node $u$ with an incident edge that can be added to the network
    \item Determine edge $(u,v) \in \bar{E}$ to add to the network
    \item Determine if $T$ belongs to the set of optimal partitions for $G' = (V(G), E(G) \cup \{(u,v)\}$
\end{enumerate}
From the computations on change in modularity in Appendix \ref{app:proofs}, we expect that choosing such an edge will decrease the gap between $Q_{P^*}$ and $Q_T$, which works towards enforcing that $T$ is a partition that maximizes modularity. We describe the procedure for unweighted networks first, then discuss how the procedure can be modified for weighted networks.

\subsubsection{Identifying Most ``Incorrectly Classified" Nodes}
Before we can choose the node that is ``most incorrectly classified", we need to first determine which nodes are incorrectly classified. As with defining the disjunctive cuts, we create an auxiliary bipartite network based on the clusters of $P^*$ and $T$, and solve the maximum weight matching problem on this network. To determine which nodes are incorrectly classified, we solve the maximum weight matching problem as described in Section \ref{sec:disj}. Let $M$ be the set of incorrectly classified nodes.

We now choose which incorrectly classified node $u \in M$ will be incident to the added edge. We do so by comparing the number of nodes that $u$ should be clustered with in $T$ but is currently not in a partition $P^*$ that currently maximizes modularity, to the number of nodes it should be clustered with. Let $C^P (u)$ is the set of nodes in the cluster of $P$ containing $u$. We order the nodes in $M$ in decreasing order, determined by the fraction
\begin{equation}
    q_u = \frac{|C^T (u) \setminus C^{P^*} (u)|}{|C^T (u)| - 1}\text{.}
\end{equation}
Unless $C^T(u) = \{ u\}$, the denominator of this term will be strictly greater than $0$, and thus the term is well defined. Brandes et al.\ \cite{brandes2007modularity} proves that, as long as there are no isolated nodes in the network, a partition that maximizes modularity will never have a cluster that is just a single node. Additionally, they show that an isolated node can be placed in any cluster (including its own) without impacting modularity. Thus, if a ground truth partition ever has a cluster that is a single node $u$, that problem will either be infeasible if $d_u >0 $, or that node can be placed in its own cluster in a partition that maximizes modularity, so it will never be incorrectly classified.

By looking at the proportion of two terms of $q_u$, we can differentiate between different size clusters. If there are two nodes $i$ and $j$ such that $|C^T (i) \setminus C^{P^*} (i)| = |C^T (j) \setminus C^{P^*} (j)|$, and $|C^T (i)| < |C^T (j)|$, then $\frac{|C^T (i) \setminus C^{P^*} (i)|}{|C^T (i) - 1|} > \frac{|C^T (j) \setminus C^{P^*} (j)|}{|C^T (j) - 1|}$. By normalizing the clusters by their size, we allow for smaller clusters to more easily distinguish themselves and be selected earlier in the heuristic. Algorithm \ref{alg:chooseHead} formalizes this procedure.

\begin{algorithm}[h!]
\caption{Identify Most Incorrectly Classified Nodes ($MostIncorrect(G,T,P^*)$}
\label{alg:chooseHead}
\begin{algorithmic}
\STATE{\textbf{Initialize:}} network $G = (V, E)$, partitions $T$, $P^*$.
\STATE{\textbf{Step 1.}} Create auxiliary bipartite network between clusters of $T$ and $P^*$, and solve the maximum weight matching problem to determine set of incorrectly classified nodes $M$.
\STATE{\textbf{Step 2.}} Order $M = \{u_1, \ldots, u_{|M|}\}$ in descending order based on $\frac{|C^T (u) \setminus C^{P^k} (u)|}{|C^T (u) - 1|}$. Return $M$.
\end{algorithmic}
\end{algorithm}

\subsubsection{Determining Beneficial Edges to Add}
Once we have determined which nodes to prioritize adding an edge to, we must then determine which edge should be added to the network. Consider $u \in M$. Our first priority is to add an edge $(u,v)$ such that $v \in C^T (u)$ and $v \notin C^{P^*} (u)$. We expect that adding such an edge to $G$ will increase the modularity of $T$ while decreasing the modularity of $P^*$. Let $N(u) = \{j \in V: j \in C^T (u)\setminus C^{P^*} (u) \land \ (u,v) \in \bar{E}\}$. Intuitively, $N(u)$ is the set of  all nodes that $u$ should be clustered with but currently is not, and an edge incident to $u$ can be added to the network. Adding an edge where $v \in N(u)$ achieves our desired goal.

If $N(u)$ is empty, then no such edge can be added to the network, and we consider the next node in $M$. If, for every $u \in M$, $N(u) = \emptyset$, no edge can be added such that it is within a cluster of $T$ and between clusters of $P^*$. Since we cannot decrease the modularity of $P^*$ without also decreasing the modularity of $T$, we additionally consider edges that increase the modularity of both $T$ and $P^*$. In some cases, such as when some cluster of $T$ is a subset of a cluster of $P^*$, the edge we add to the network will increase the modularity of $T$ more than it increases the modularity of $P^*$. 

Instead of considering $N(u)$, we consider $\Tilde{N}(u) = \{j \in V: j \in C^T (u) \land \ (u,v) \in \bar{E}\}$. This allows us to add edges that are both within $C^T(u)$ and $C^{P^*}$, relaxing our restriction to include edges we expect to also increase the modularity of $P^*$. If, for every $u \in M$, $\Tilde{N}(u) = \emptyset$, then no incorrectly classified node can have an incident edge added to the network. As such, we expect that every edge we can add to the network will \emph{decrease} the modularity of $T$; we use this as termination criterion under which the problem may be infeasible.

Suppose that, for some $u \in M$, there exists a node $v \in N(u)$ (or $v \in \Tilde{N}(u)$). Thus, an edge incident to $u$ can be added to the network that we expect will increase the modularity of $T$. We now must choose which of these edges to add to the network. Recall that the change in modularity by adding an edge to the network is independent of the nodes incident to the edge, and only dependent on the cluster those nodes belong to. Thus, for a fixed partition, all of the considered edges will have the same impact on the modularity of the partition. This motivates us to consider how the edge we choose to add will impact which partition is optimal in the network resulting from adding that edge.

We want to add an edge $(u,v)$ such that, when we find an optimal partition $P'$ of the resulting network $G' = (V, E \cup \{(u,v)\})$, $u$ is in a different cluster of $P'$ than $P^*$, moving $u$ to be clustered with a node it should be classified with. However, if $v$ is not sufficiently connected to the cluster it belongs to, adding $(u,v)$ could result in the opposite change of clustering, moving $v$ into a different cluster in the resulting network. To avoid such a scenario, we measure $v$'s contribution to the change in modularity by adding $(u,v)$, which we define to be $\delta_v$. Multiplying by $(2m)^2 (2m+2)^2$, we compute this as follows:

\begin{align*}
    \delta_v & = (2m)^2(2m+2) A_{vv} - (2m)^2 (d_v+1)^2 - (2m)(2m+2)^2 A_{vv} + (2m+2)^2 d_v^2 \\
    & + 2\left((2m)^2(2m+2) A_{uv} + (2m)^2(2m+2) - (2m)^2(d_u+1)(d_v+1)\right)\\ & - 2\left((2m)(2m+2)^2 A_{uv} + (2m+2)^2 d_u d_v \right)\\
    & + 2\left(\sum_{u,v \ne y \in C} \left((2m)^2(2m+2)A_{vy} - (2m)^2(d_v+1)d_y - (2m)(2m+2)^2 A_{vy} + (2m+2)^2 d_v d_y\right) \right) \\
    &\\
    & = (-8m^2 - 8m)A_{vv} - 8m^2 d_v - 4m^2 +8m d^2_v + 4d^2_v\\
    & + 2\left((-8m^2 - 8m)A_{uv} + 8m^3 + 8m^2 - (4m^2)(d_u d_v + d_u + d_v +1) + (4m^2 + 8m + 4) d_u d_v \right)\\
    & + 2\left(\sum_{u,v \ne y \in C} \left( (-8m^2 -8m)A_{vy} - 4m^2(d_v d_y + d_y) + (4m^2 + 8m +4) d_v d_y  \right)\right)\\
    &\\
    & + (-8m^2 - 8m)A_{vv} + (8m +4) d^2_v - 8m^2 d_v - 4m^2\\
    & + 2\left((-8m^2 - 8m)A_{uv}  + (8m +4) d_u d_v + 8m^3 + 4m^2 - 4m^2 d_u -4m^2 d_v \right)\\
    & + 2\left(\sum_{u,v \ne y \in C} \left( (-8m^2 -8m)A_{vy} + (8m+ 4)d_v d_y - 4m^2 dy  \right)\right)\\
    &\\
    & = 16m^3 + 4m^2 + (-8m^2 - 8m)A_{vv} - 16m^2 d_v \\ &+  2\sum_{y \ne v, y \in C} \left((-8m^2 - 8m) A_{vy} + (8m+ 4)d_v d_y - 4m^2 d_y\right)
\end{align*}

We drop the $16m^3 + 4m^2$ term, since that will be constant across all choices of $v$, reducing our rule to 

\begin{equation*}
    \delta_{v} = (-8m^2 - 8m)A_{vv} - 16m^2 d_v +  2\sum_{y \ne v, y \in C} \left((-8m^2 - 8m) A_{vy} + (8m+ 4)d_v d_y - 4m^2 d_y\right) \text{.}
\end{equation*}

 We choose to take the node $v^* = \text{arg}\min_{v \in N(u)} \delta_v$, as a smaller local change in modularity indicates that node $v$ is better connected within its cluster. We formalize this procedure in Algorithm \ref{alg:chooseTail}.

\begin{algorithm}[h!]
\caption{Choosing a Beneficial Edge ($ChooseEdge(G,T,M)$}
\label{alg:chooseTail}
\begin{algorithmic}
\STATE{\textbf{Initialize:}} network $G = (V, E)$, partition $T$, ordered set of incorrectly classified nodes $M = \{u_1, \ldots, u_{|M|}\}$, $chosen=false$, $l=0$.
\WHILE{$chosen = false$}
\STATE{\textbf{Step 1.}} Set $l \leftarrow l+1$.
 \IF{$l \le |M|$}
 \STATE{\textbf{Step 2a.}} Identify candidate nodes $N(u_l) =\{j \in V: C^T (u_l) \setminus C^{P^k} (u_l) \land (u_l,v) \in \bar{E}\}$.
 \IF{$N(u_l) = \emptyset$}
 \STATE{Return to Step 1.}
 \ELSE
 \STATE{\textbf{Step 3a.}} Set $u = u_l$.
 \STATE{\textbf{Step 3b.}} Compute $\delta_j$ for all $j \in N(u)$.
 \STATE{\textbf{Step 3c.}} Choose $j \in \text{arg}\min_{i \in N(u)} \delta_i$. Return $(u,v)$.
 \ENDIF
 \ELSIF{$l > |M|$ and $l \le 2|M|$}
 \STATE{\textbf{Step 2b}} Set $\Tilde{l} = l -|M|$
 \STATE{\textbf{Step 2c.}} Identify candidate nodes $\Tilde{N}(u_{\Tilde{l}}) =\{j \in V: C^T (u_{\Tilde{l}}) \land (u_{\Tilde{l}},v) \in \bar{E}\}$.
 \IF{$\Tilde{N}(u_{\Tilde{l}}) = \emptyset$}
 \STATE{Return to Step 1.}
 \ELSE
 \STATE{\textbf{Step 4a.}} Set $u = u_{\Tilde{l}}$
 \STATE{\textbf{Step 4b.}} Compute $\delta_j$ for all $j \in \Tilde{N}(u)$.
 \STATE{\textbf{Step 4c.}} Choose $v \in \text{arg}\min_{i \in \Tilde{N}(u)} \delta_i$. Return $(u,v)$.
 \ENDIF
 \ELSE
 \STATE{Terminate, problem is infeasible.}
 \ENDIF
 \ENDWHILE
\end{algorithmic}
\end{algorithm}

We note that our choice is closely related to the degree of each candidate node. Intuitively, this aligns with Theorem \ref{thrm:star}. By adding edges to the highest degree node within a cluster, we work towards creating a star sub-network in the cluster. For incorrectly classified nodes connected to many nodes in other clusters, we need to add more edges than just establishing a base framework.

Once we have added an edge to the network to create $G'$, we now need to determine if $T$ belongs to the set of partitions that maximize the modularity of $G'$. We do so by finding a partition $P'$ that maximizes the modularity of $G'$, and computing the modularity of $T$. If $Q_T = Q_{P'}$, then $T$ belongs to the set of optimal partitions, and we can terminate. Otherwise, we repeat the procedure. We formalize the full procedure in Algorithm \ref{alg:alg2}.

\begin{algorithm}[h!]
\caption{Heuristic Algorithm for Edge Addition for Community Optimization}
\label{alg:alg2}
\begin{algorithmic}
\STATE{\textbf{Initialize:}} initial network $G^1 = (V, E)$, set of edges added $E' = \emptyset$, ground truth partition $T$, iteration counter $k=1$.
 \WHILE{not converged}
 \STATE{\textbf{Step 1.}} Find partition $P^*$ that maximizes modularity of $G$.
 \IF{$Q_{T}(G^k) = Q_{P^*}(G^k)$}
 \STATE{Terminate; $T$ belongs to the set of optimal partitions. Return $E'$}
 \ELSE
 \STATE{\textbf{Step 2.}} Set M = $MostIncorrect(G^k,T,P^*)$.
 \STATE{\textbf{Step 3.}} Set $(u,v) = ChooseEdge(G^k,T,M)$.
 \STATE{\textbf{Step 4.}} Set $G^{k+1} = (V(G^k), E(G^k) \cup (u,v))$, $E' \leftarrow E' \cup (u,v)$, $k \leftarrow k + 1$.
 \ENDIF
\ENDWHILE
\end{algorithmic}
\end{algorithm}

\begin{figure}[ht]
\centering
\begin{subfigure}{.32\textwidth}
  \centering
  \includegraphics[width=\linewidth]{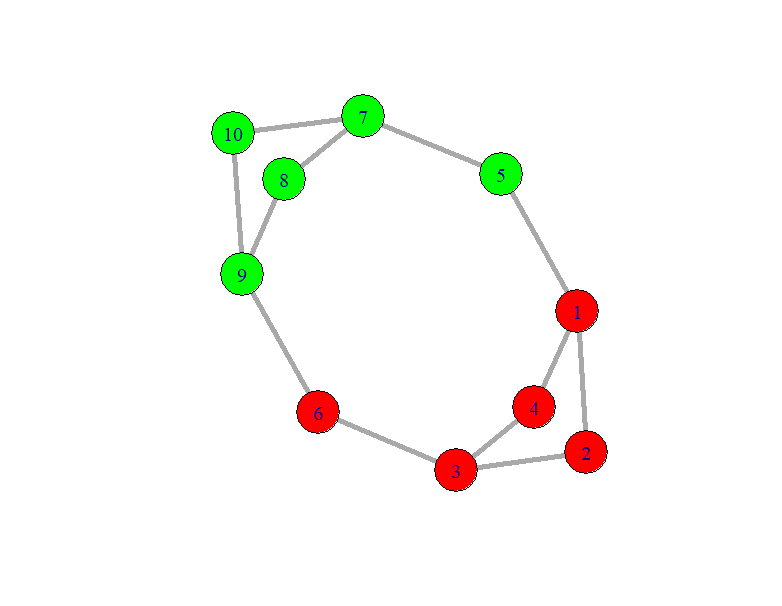}
  \caption{Optimal Partition}
  \label{fig:rev1}
\end{subfigure}%
\begin{subfigure}{.32\textwidth}
  \centering
  \includegraphics[width=\linewidth]{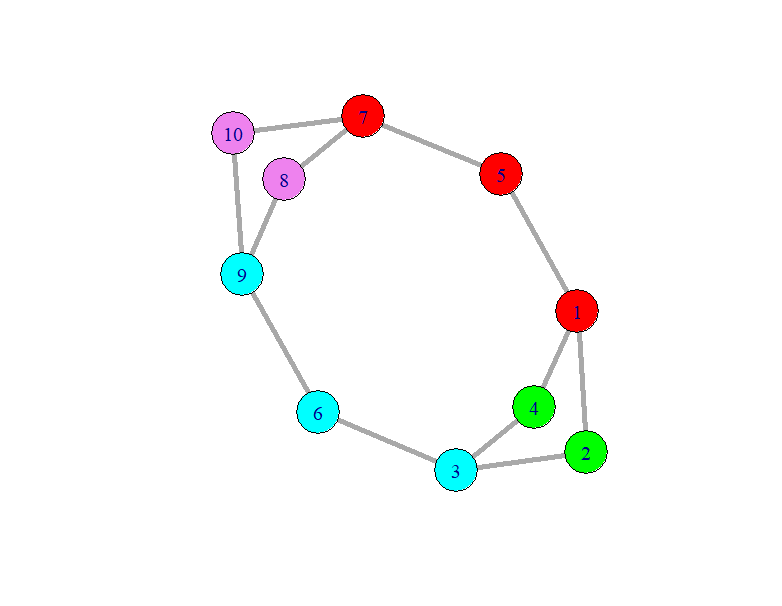}
  \caption{Ground Truth}
  \label{fig:rev2}
\end{subfigure}
\begin{subfigure}{.32\textwidth}
  \centering
  \includegraphics[width=\linewidth]{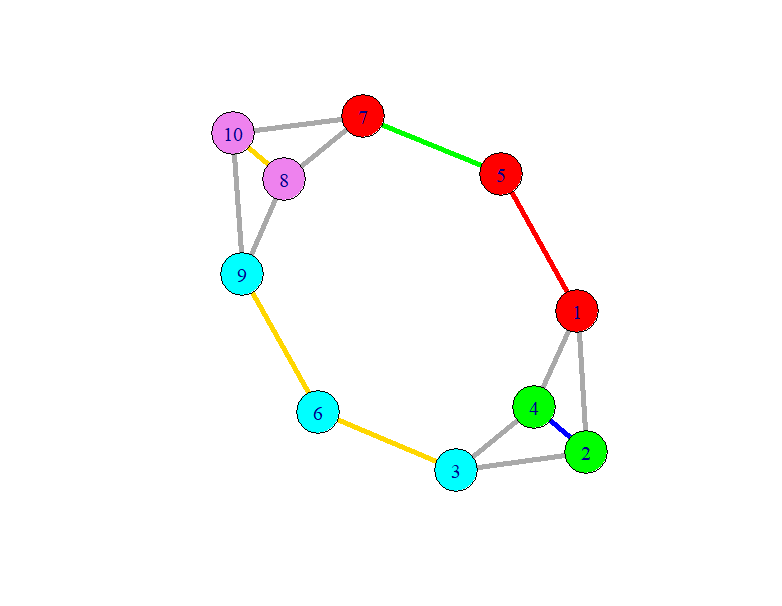}
  \caption{Ground Truth with Added Weights}
  \label{fig:rev3}
\end{subfigure}%
\caption{Example network where an edge needs additional weight added after first visit by the heuristic approach}
\label{revisit}
\end{figure}

Extending this heuristic to weighted networks is quite simple. While computations of $m$ and $d_i$ change, the above computation remains the same. We additionally allow for weight to be added to edges already existing in the network, removing the condition $A_{uv}=0$ from the definition of $N(u)$ and $\Tilde{N}(u)$. In a weighted network, we note that we could have instead added weight to an edge until the partition changes. Even if we were to do so, this does not guarantee that an incorrectly classified node will be correctly classified in future iterations nor that an edge will not be re-visited by the heuristic in future iterations, as demonstrated in the example in Figure \ref{revisit}. In this example, each edge has weight $1$, and the weight limit for each edge is $10$.

In this network, weight is added to the colored edges in Figure \ref{fig:rev3}. In the first iteration, a unit of weight is added to the red edge, $(1, 5)$, causing nodes $1$ and $5$ to be clustered together. The second iteration adds the blue edge, $(2, 4)$. The third and fourth iterations add a unit of weight to the green edge $(5, 7)$. The addition of this weight causes nodes $5$ and $7$ to be clustered together, but separates node $1$ from node $5$. Our algorithm next adds another weight to the red edge. At the beginning, only one unit weight is needed for node $1$ to be classified correctly. However, after more iterations, node $1$ is no longer classified correctly, and an edge needs to be revisited.

\subsection{Heuristic Algorithm Edge Removal}
When considering model \eqref{model2}, instead of selecting which edge to include from the edge set, we turn our attention to removing edges from the network. The idea behind this procedure is quite intuitive as well; order the edges in the network, then, one by one, check if removing the $i^{th}$ edge results in a change in an optimal partition. If the optimal partition does not change, then that edge can be removed from the network. Note that, if the $i^{th}$ edge was removed, that edge could have been critical for changing the optimal partition when removing any of the first $i-1$ edges. Thus, if any single edge is removed, we must check the full edge list again to determine if more edges can be removed. We repeat this procedure until it determines that removing any single edge in the network results in a change in an optimal partition. We formalize this approach in Algorithm \ref{alg:alg3}.

\begin{algorithm}
\caption{Heuristic Algorithm for Edge Removal for Community Preservation}
\label{alg:alg3}
\begin{algorithmic}
\STATE{\textbf{Initialize:}} initial network $G^1 = (V, E)$, set of edges removed $E' = \emptyset$, optimal partition $T$, $check = true$, iteration counter $k=1$.
\WHILE{$check = true$}
\STATE{\textbf{Step 1.}} Set $check = false$, $I = \emptyset$.
\STATE{\textbf{Step 2.}} Order edges $E(G^k) = \{e_1, \ldots, e_{|E(G^k)|}\}$.
\FOR{$i \in \{1, \ldots, |E(G^k)|\}$}
\STATE{\textbf{Step 3.}} Create network $G' = (V(G^k), E(G^K)\setminus(I \cup \{e_i\}))$.
\STATE{\textbf{Step 4.}} Find an optimal partition $P^*$ of $G'$.
\IF{$P^* = T$}
\STATE{\textbf{Step 4a.}} Set $I \leftarrow I \cup \{e_i\}$, $check = true$.
\ENDIF
\ENDFOR
\STATE{\textbf{Step 5.}} Create network $G^{k+1} = (V(G^k), E(G^k)\setminus I)$.
\STATE{\textbf{Step 6.}} Set $E' \leftarrow E' \cup I$, $k \leftarrow k+1$.
\ENDWHILE
\end{algorithmic}
\end{algorithm}

We test the following rules for ordering the edges:
\begin{enumerate}
    \item a prespecified order,
    \item a random permutation,
    \item a fixed order based on contribution to modularity,
    \item a dynamically updated order based on contribution to modularity.
\end{enumerate}
A weakness of the first rule is that the prespecified order may choose edges whose removal may cause fewer edges to be removed in subsequent iterations. The second rule randomizes the order in which the edges are processed in each iteration. This provides a workaround for the weakness of the first rule, but the quality of the solution obtained using this rule is determined by the orders randomly generated.

The third and fourth rules use an edge's contribution to modularity to determine order. We order the edges in decreasing order of their values in the modularity matrix, $M_{ij} = A_{ij} - \frac{d_i d_j}{2m}$. Intuitively, edges with a large value of $M_{ij}$ have small products of degrees, meaning this edge has a low probability to be generated in the corresponding random network. We suspect that these edges are unlikely to be impactful in determining an optimal partition, making them good candidates for removal. For our third rule, we choose to pre-compute this ordering and use it in each iteration, as with the first rule. For the fourth rule, we update this ordering after each iteration. Since it is possible that the removal of certain edges can cause changes in the importance of other edges to an optimal partition, the updates may result in more edges removed.

Since Algorithm \ref{alg:alg3} ensures that each subsequent network has the same optimal partition as the original network, we can improve the solution found by Algorithm \ref{alg:alg2} by taking the output solution $G'$ as the input for Algorithm \ref{alg:alg3}. This removes edges that were added, but later become redundant due to the addition of other edges. For this addition as post-processing, we simply choose to process the edges in the order in which they were added. Extending this algorithm to weighted networks is also straightforward. Instead of removing one edge at a time, we remove a unit of weight on one edge at a time.

\section{Computational Results}
We test our methods on multiple networks, some of which are popular in community detection literature and some popular in illicit network analysis. We retrieved this data from the CASOS Public Datasets \cite{casos} and Pajek Datasets \cite{pajek}. Table \ref{tbl:tests} reports the networks we use for our experiments organized by the number of edges in the network. The top four networks are well known in the field of community detection, while the bottom six are known illicit networks. For the Italian Gangs, London Gangs, and Montreal Gangs data sets, which have disconnected components, we delete the nodes that are not in the largest connected component. While this is not a necessary modification, we do so to simplify the networks. We model the integer programming problem in AMPL with CPLEX 20.1 \cite{studio2017cplex} as the solver. Experiments are conducted on a laptop with an Intel\textsuperscript{\textregistered} Core\textsuperscript{TM} i5-8250 CPU @ 1.6 GHz - 1.8 GHz and 16 GB RAM running Windows 10. We set a time limit of $7200$ seconds. We develop the heuristics in R using the iGraph package \cite{igraph}. Due to integer programming limitations, we additionally restrict the set of edges considered for addition in these experiments. To ensure feasibility, we start with the set of edges identified by the heuristics with post-processing and without edge restrictions as the initial set of allowable edges. We then generate the list of all pairs of nodes with shortest path length $2$ between them, and randomly select from that list to add to the set of allowable edges. We consider $100$ edges for each network.

\begin{table}[H]
 \begin{center}
 \caption{Size of test networks}
  \begin{tabular}{|c|c|c|} 
  \hline
  Network & $|V|$& $|E|$ \\ [0.5ex] 
  \hline\hline
  Sawmill \cite{michael1997modeling} & 36 & 62\\
  \hline
  Dolphins small \cite{lusseau2003bottlenose} & 40 & 70 \\
  \hline
  Karate \cite{zachary} & 34 & 78  \\ 
  \hline 
  Les Mis \cite{knuth1993stanford} & 77 & 254 \\
  \hline \hline
  Ciel \cite{caviar6} & 25 & 35\\
  \hline
  Caviar6 \cite{caviar6} & 27 & 47 \\
  \hline
  Rhodes \cite{rhodes2007social} & 22 & 66 \\
  \hline
  Montreal Gangs \cite{descormiers2011alliances} & 29 & 75 \\
  \hline
  Italian Gangs \cite{casos} & 65 & 113\\
  \hline
  London Gangs \cite{grund2015ethnic} & 48 & 133 \\
  \hline
  \end{tabular}
 \label{tbl:tests}
 \end{center}
\end{table}

\subsection{Results for Edge Addition to Unweighted Networks}
\begin{figure}
        \begin{subfigure}{0.49\linewidth}
            \centering
            \captionsetup{justification=centering}
            \includegraphics[width=\textwidth]{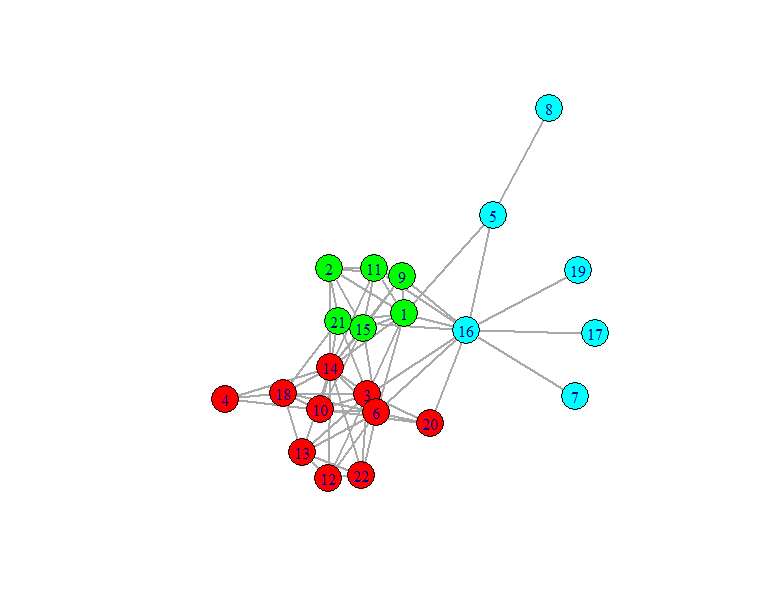}
            \caption{Optimal Partition}
            \label{fig:cav6oc}
        \end{subfigure}
        \begin{subfigure}{0.49\linewidth}
            \centering
            \captionsetup{justification=centering}
            \includegraphics[width=\textwidth]{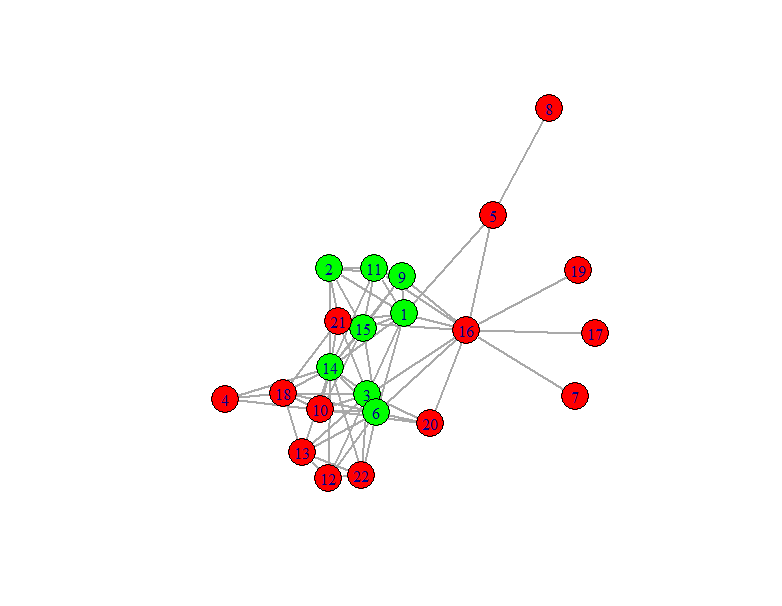}
            \caption{Ground Truth}
            \label{fig:cav6gt}
        \end{subfigure}
        \caption{Optimal partition and ground truth community structure of Rhodes Network}
        \label{rhodesplot}
\end{figure}

\begin{figure}
        \begin{subfigure}{0.49\linewidth}
            \centering
            \captionsetup{justification=centering}
            \includegraphics[width=\textwidth]{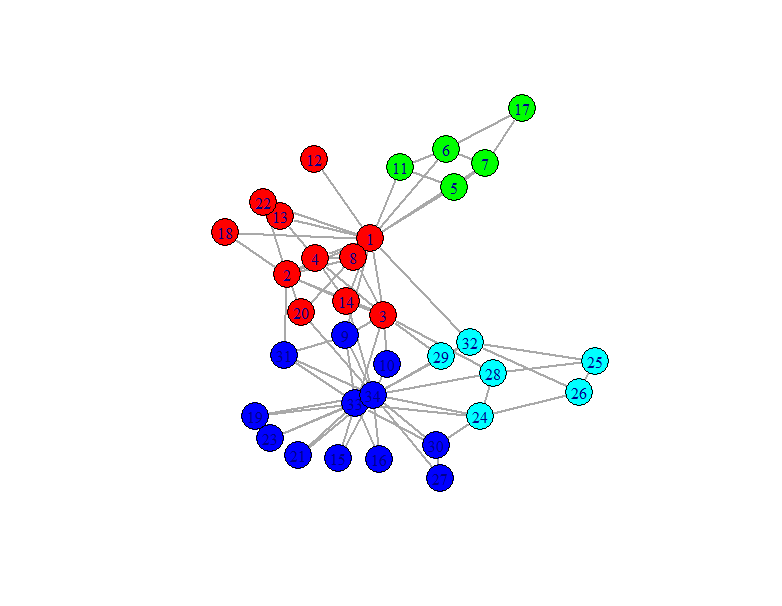}
            \caption{Optimal Partition}
            \label{fig:karateoc}
        \end{subfigure}
        \begin{subfigure}{0.49\linewidth}
            \centering
            \captionsetup{justification=centering}
            \includegraphics[width=\textwidth]{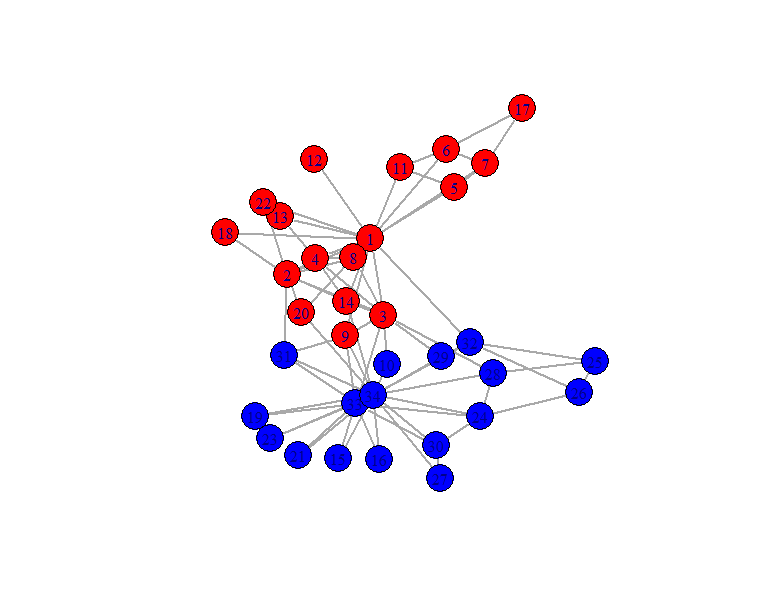}
            \caption{Ground Truth}
            \label{fig:karategt}
        \end{subfigure}
        \caption{Optimal partition and ground truth community structure of Karate Club Network}
        \label{karateplot}
\end{figure}

For networks with a known community structure, we use it as the ground truth. For the \emph{Les Mis\'{e}rables} (Les Mis), London Gangs, and Italians Gangs data sets, we use the greedy clustering method \cite{newman2004fast} to identify a partition to be the ground truth. For the Ciel data set, we use the Louvain method \cite{blondel2008fast} to identify a partition to be ground truth. Using the greedy clustering method resulted in partitions that had more differences in clusters than the Louvain method, except with the Ciel data set, where the greedy clustering method outputs an optimal partition. For networks that are originally weighted, we remove edge weights and experiment on their unweighted versions. Figures \ref{rhodesplot} and \ref{karateplot} demonstrate the differences in an optimal partition and the ground truth community structure for example networks.

Table \ref{tbl:ur_e} reports the initial number of incorrectly classified nodes as well as the number of edges added to each network to reach the desired communities. Table \ref{tbl:ur_t} reports the run time of each method. We denote the integer program without disjunctive cuts and extra partitions as IP, and the integer program with disjunctive cuts and additional partitions as IP+, and the number of partitions visited by our algorithm as nParts. Entries marked with * did not converge to the optimal solution within the time limit, and the lower bound on the optimal solution is reported. For instances where the integer programming formulation did not solve the problem to optimality, Table \ref{tab:optgap} reports the difference in modularity between the ground truth partition and an optimal partition after adding in the edges determined as the best known lower bound.

\begin{table}[H]
 \begin{center}
 \caption{Comparison of Number of Edges Added to Tested Networks with Edge Restrictions}
 \resizebox{\textwidth}{!}{
  \begin{tabular}{|c|c|c|c|c|c|c|} 
  \hline
  Network & Init \# Misclassified &\#Edges Added (IP) & \#Edges Added (IP+)& \#Edges Added (Alg4) & \#Edges Added(Alg4\&5)\\ [0.5ex] 
  \hline\hline
  Sawmill & 10 & 9 & 9 & 16 & 13 \\
  \hline
  Dolphins small & 17 & 24* & 29* & 34 & 30  \\ 
  \hline
  Karate & 12 & 8 & 8 & 11 & 9\\ 
  \hline
  Les Mis & 16 & 18 & 18 & 21 & 21 \\
  \hline\hline
  Ciel & 3 & 1 & 1 & 1 & 1 \\ 
  \hline
  Caviar6 & 3 & 11 & 11 & 12 & 12\\
  \hline
  Rhodes & 10 & 18* & 18* & 34 & 34 \\ 
  \hline
  Montreal Gangs & 16 & 18* & 18* & 45 & 42 \\ 
  \hline
  Italian Gangs & 21 & 2 & 2 & 4 & 4 \\
  \hline
  London Gangs & 8 & 5 & 5 & 7 & 7 \\
  \hline
 \end{tabular}}
 \label{tbl:ur_e}
 \end{center}
\end{table}

\begin{table}[H]
 \begin{center}
 \caption{Run Time (in seconds) on Tested Networks with Edge Restrictions}
  \begin{tabular}{|c|c|c|c|c|} 
  \hline
  Network & IP (nParts) & IP+ (nParts) &  Alg4 & Alg4\&5\\ [0.5ex] 
  \hline\hline
  Sawmill & 309.08 (14) & 3073.24 (8) & 4.83 & 14.25 \\
  \hline
  Dolphins small & 7200* (8) & 7200* (9) & 26.47 & 258.66 \\ 
  \hline
  Karate & 537.84 (16) & 72.63 (7) & 3.80 & 13.82\\ 
  \hline
  Les Mis & 507.69 (9) & 255.67 (6) & 690.38 & 1356.53 \\
  \hline\hline
  Ciel & 1.67 (1) & 1.67 (1) & 0.09 & 0.13 \\ 
  \hline
  Caviar6 & 126.81 (10) & 36.53 (5) & 4.95 & 10.68 \\
  \hline
  Rhodes & 7200* (9) & 7200* (6) & 4.69 & 7.31 \\ 
  \hline
  Montreal Gangs & 7200* (2) & 7200* (2) & 627.23 & 1256.49 \\ 
  \hline
  Italian Gangs & 33.34 (4) & 41.98 (3) & 42.02 & 66.89 \\
  \hline
  London Gangs & 66.17 (8) & 43.11 (4) & 26.23 & 42.14 \\
  \hline
 \end{tabular}
 \label{tbl:ur_t}
 \end{center}
\end{table}

\begin{table}[H]
    \centering
    \caption{Optimality Gap in Modularity for Unsolved Instances}
    \resizebox{\textwidth}{!}{
    \begin{tabular}{|c|c|c|c|c|c|c|}
        \hline
        Network & $Q_{P^*}(z^*)$ (IP) & $Q_{T}(z^*)$ (IP) & $Q_{P^*}(z^*)-Q_{T}(z^*)$ (IP) & $Q_{P^*}(z^*)$ (IP+) & $Q_{T}(z^*)$ (IP+) & $Q_{P^*}(z^*)-Q_{T}(z^*)$ (IP+)\\
        \hline
         Dolphins small & 0.4845518 & 0.4586351 & 0.0259167 & 0.4759208 & 0.4502092 & 0.02571166 \\
         \hline\hline
         Rhodes & 0.1748866 & 0.154691 & 0.02019558 & 0.1868622 & 0.154691 & 0.0321712\\
         \hline
         Montreal Gangs & 0.2429761 & 0.128801 & 0.114175 & 0.2099087 & 0.1315759 & 0.07833276\\
         \hline
    \end{tabular}}
    \label{tab:optgap}
\end{table}

Table \ref{tbl:ur_e} shows that the integer programming formulation is able to find a relatively small number of edges compared to the density of the network in the case of the Karate club and Ciel networks. For some networks, such as Caviar6 and Montreal Street Gangs, the layout of the network requires edges to be added to distinguish a smaller cluster from a larger cluster that can relatively easily absorb it, as well as a significant number of edges that connect nodes that are clustered together in the ground truth, but do not have any edges in the between them in the network. However, for the other networks, the integer programming formulation is not able to solve the problem within the time limit. This is due to the edge-anonymity property of modularity, as that unintentionally creates a lot of symmetry in the branch-and-bound procedure for the master problem. Integer programming problems with symmetry are particularly difficult to solve \cite{ostrowski2011orbital}. These symmetries prevented the branch-and-bound procedure from meaningfully increasing the lower bound on the objective value when fathoming nodes in the branch-and-bound tree. Additionally, as the number of edges being included increases, the number of symmetric solutions increases exponentially. We observed that, in early iterations and in subsequent iterations where the master problem objective value stays the same, the master problem is solved to optimality relatively quickly. However, in later iterations when the number of edges needed must be larger than in the previous iteration, the master problem may not necessarily solve to optimality due to the number of branches that need to be explored. We suspect that, in each iteration of the master problem, the integer program identifies the optimal solution with respect to the set of previously explored partitions, but does not complete the branching procedure to verify that the identified solution is indeed optimal. Appendix \ref{app:time} presents the comparison in solve time between the master and subproblems, with the default symmetry breaking setting and maximum symmetry breaking setting.

We note that this issue is not solely based on the size of the network. The integer programming framework was able to solve the problem on the Les Mis, Italian Gangs, and London Gangs networks, but not on the Dolphins, Rhodes or Montreal Gangs networks. The former three networks all have more nodes and edges than the latter three networks. Additionally, the former three networks all had ground truth communities determined by using the greedy clustering procedure. This indicates that the difficulty of the problem is more dependent on the differences between the ground truth community and the initial optimal community structure than the size of the network.

Including disjunctive cuts and additional partitions improves the solve time of the integer programming formulation for the Karate club network, and only hinders the solve time of the Sawmill network. This indicates that the extra information generated may have prevented a useful partition that was visited by the base integer program from being visited by the augmented procedure. For the Dolphins network, the best known lower bound increases from $24$ to $29$, drastically improving the result, and improving the known quality of the heuristic solution. For Montreal Gangs and Rhodes, no improvements are made by including the disjunctive cuts and additional partitions. This indicates that the edge-anonymity caused the solver to stall early on in this procedure. Increasing the run time to $36000$ seconds on these instances shows that little-to-no progress is made in improving these lower bounds for either procedure. This demonstrates that problem difficulty is not only dependent on the size of the network, but the density of the network and difference between the ground truth community structure and initial optimal partition.

The heuristic approach is able to find reasonably high quality solutions as compared to the integer program optimal solutions before post-processing is implemented. With post-processing, we are able to reduce the number of edges needed in some networks. Furthermore, where the integer program may not solve the problem in the time limit, the heuristic approach is able to find a solution in under an hour in all instances. When the integer program does solve the problem within the time limit, the heuristic approach is able to solve the problem significantly faster. We note that, since some of the allowable edges are determined by the heuristic without edge restrictions, adding the same edge restrictions to the heuristic will result in the same solution. However, allowing any edge currently not in the network to be added to the integer programming formulation would drastically increase the run time.

We observe that when the ground truth communities are relatively similar to the initial optimal community structure, fewer edges are needed. Additionally, if the ground truth community structure has clear boundaries between the communities, fewer edges are needed. Figures \ref{rhodesplot} and \ref{karateplot} help distinguish this difference. In Figure \ref{fig:karategt}, there are very few edges between the two clusters, and there are very few nodes that have edges that cross between clusters. This results in fewer edges being needed to properly cluster these boundary nodes. However, in Figure \ref{fig:cav6gt}, the distinction between the clusters is significantly less clear, resulting in a need for each cluster to have more edges added to properly distinguish them. Another example of this is the difference between the Caviar6 and Ciel networks. While both have only three initially incorrectly classified nodes, the Caviar6 network has a cluster that is mostly absorbed into another cluster, where the boundaries between the clusters in the Ciel network are more distinct. This causes a need for more edges in the Caviar6 network to properly distinguish this cluster.

\subsection{Results for Edge Weight Addition to Weighted Networks}
We also test our methods on weighted versions of each network used in our unweighted test set. Networks that are already weighted have an upper bound on the weight of edge set to the weight of the maximum weight edge. For networks that are unweighted, we randomly generate weights between $1$ and $15$ for each edge, with a maximum edge weight of $15$ allowed. We use the unweighted optimal partition as the ground truth, and verify that an optimal partition for the weighted network is not the same as that of the unweighted network. We note that, unlike with unweighted networks, we cannot compare the heuristic results directly against the integer programming results. The integer program cannot find the optimal solution due to the inaccuracy in the McCormick inequalities. Table \ref{tbl:weights} reports the weights on networks used for experiments, using the same ordering as above.

\begin{table}[h]
 \begin{center}
 \caption{Total Weights of Edge Sets of Test Networks}
  \begin{tabular}{|c|c|} 
  \hline
  Network & Total Weight \\ [0.5ex] 
  \hline\hline
  Sawmill & 481\\
  \hline
  Dolphins small & 544\\
  \hline
  Karate & 654 \\ 
  \hline
  Les Mis & 2030\\
  \hline\hline
  Ciel & 198 \\
  \hline
  Caviar6 & 668 \\
  \hline
  Rhodes & 524\\
  \hline
  Montreal Gangs & 621\\
  \hline
  Italian Gangs & 915\\
  \hline
  London Gangs & 980\\
  \hline
 \end{tabular}
 \label{tbl:weights}
 \end{center}
\end{table}

As before, we restrict the set of edges to which weight can be added to in each network. We restrict this to be the set of currently existing edges, enforcing that no new edges can be added. Additionally, we restrict the total weight that an edge can have. Table \ref{tbl:wr_e} reports the initial number of incorrectly classified nodes and the amount of weight added to each network. Table \ref{tbl:wr_t} reports the run time of each method.

\begin{table}[H]
 \begin{center}
 \caption{Comparison of Amount of Weight Added to Tested Networks with Edge Restrictions}
 \resizebox{\textwidth}{!}{
  \begin{tabular}{|c|c|c|c|} 
  \hline
  Network & Init. \#Misclassified & Weight Added (Alg4) & Weight Added(Alg2\&3)\\ [0.5ex] 
  \hline\hline
  Sawmill & 16 & 30 & 26 \\
  \hline
  Dolphins small & 8 & 61 & 58 \\
  \hline
  Karate & 13 & 45 & 44 \\ 
  \hline
  Les Mis & 11 & 48 & 47* \\
  \hline\hline
  Ciel & 3 & 12 & 12 \\
  \hline
  Caviar6 & 7 & 115 & 113 \\
  \hline
  Rhodes & 6 & 38 & 31 \\
  \hline
  Montreal Gangs & 11 & 45 & 43 \\
  \hline
  Italian Gangs & 16 & 30 & 30 \\
  \hline
  London Gangs & 17 & Infeasible* & Infeasible* \\
  \hline
 \end{tabular}}
 \label{tbl:wr_e}
 \end{center}
\end{table}

\begin{table}[H]
 \begin{center}
 \caption{Run Time (in seconds) on Tested Networks with Edge Restrictions}
  \begin{tabular}{|c|c|c|} 
  \hline
  Network & Alg2 & Alg2\&3\\ [0.5ex] 
  \hline\hline
  Sawmill & 11.02 & 48.58\\
  \hline
  Dolphins small & 20.79 & 68.19 \\
  \hline
  Karate & 46.15 & 120.11\\ 
  \hline
  Les Mis & 2243.14 & 7200* \\
  \hline \hline
  Ciel & 0.51 & 0.96 \\
  \hline
  Caviar6 & 17.42 & 60.38 \\
  \hline
  Rhodes & 2.64 & 7.37 \\
  \hline
  Montreal Gangs & 51.06 & 169.83 \\
  \hline
  Italian Gangs & 158.94 & 367.81 \\
  \hline
  London Gangs & 6073.99* & 6073.99* \\
  \hline
 \end{tabular}
 \label{tbl:wr_t}
 \end{center}
\end{table}

For all networks except Caviar6 and London Gangs, we are able to add less than $12\%$ of the initial total weight to the network in order to return to the desired clustering. With Caviar6, we see that this percentage is much higher, at $17\%$ of the initial total weight. This is likely due to the large disparity in distribution of edge weights; there are a few nodes with a lot of weight on incident edges, but many nodes with a small amount of weight on incident edges. Being able to correctly classify these nodes with a small amount of weight on incident edges takes much more weight. Similarly to the unweighted case, the post-processing is able to reduce the total weight added by a small amount, but this decrease is not as significant as in the unweighted case. In the case of London Gangs, the heuristic determines the problem is infeasible with edge restrictions. The weights on this network were randomly generated, so we know the heuristic is incorrect in this determination; by adding weights such that every edge in the network has the same weight, the optimal clustering will be the same as in the unweighted network. This highlights that adding weights in a greedy fashion may prevent the problem from being solved correctly.

We next report results on weighted networks without edge restrictions. Table \ref{tbl:wu_e} reports the initial number of incorrectly classified nodes, as well as the number of edges added to each network to reach the desired communities. Table \ref{tbl:wu_t} reports the run time of each method.

\begin{table}[h]
 \begin{center}
 \caption{Comparison of Amount of Weight Added to Tested Networks without Edge Restrictions}
 \resizebox{\textwidth}{!}{
  \begin{tabular}{|c|c|c|c|} 
  \hline
  Network & Init. \#Misclassified & Weight Added (Alg2) & Weight Added(Alg2\&3)\\ [0.5ex] 
  \hline\hline
  Sawmill & 16 & 26 & 22 \\
  \hline
  Dolphins small & 8 & 62 & 58 \\
  \hline
  Karate & 13 & 45 & 44 \\ 
  \hline
  Les Mis & 11 & 48 & 47* \\ 
  \hline\hline
  Ciel & 3 & 12 & 12 \\
  \hline
  Caviar6 & 7 & 115 & 113 \\
  \hline
  Rhodes & 6 & 28 & 26 \\
  \hline
  Montreal Gangs & 11 & 48 & 46 \\
  \hline
  Italian Gangs & 16 & 31 & 30 \\ 
  \hline
  London Gangs & 17 & 151 & 134 \\ 
  \hline
 \end{tabular}}
 \label{tbl:wu_e}
 \end{center}
\end{table}

\begin{table}[h]
 \begin{center}
 \caption{Run Time (in seconds) on Tested Networks without Edge Restrictions}
  \begin{tabular}{|c|c|c|} 
  \hline
  Network & Alg2 & Alg2\&3 \\ [0.5ex] 
  \hline\hline
  Sawmill & 9.67 & 36.08\\
  \hline
  Dolphins small & 21.22 & 52.27 \\
  \hline
  Karate & 50.97 & 145.72 \\ 
  \hline
  Les Mis & 2482.99 & 7200* \\ 
  \hline\hline
  Ciel & 0.50 & 0.96 \\
  \hline
  Caviar6 & 17.57 & 60.15 \\
  \hline
  Rhodes & 1.72 & 6.93 \\
  \hline
  Montreal Gangs & 34.32 & 90.14 \\
  \hline
  Italian Gangs & 187.92 & 762.48 \\ 
  \hline
  London Gangs & 4643.55 & 7200* \\ 
  \hline
 \end{tabular}
 \label{tbl:wu_t}
 \end{center}
\end{table}

In this case, removing edge restrictions results in similar overall weight being added to the networks, and has run times comparable to that of the restricted edge case. An interesting result is that less weight needs to be added to the Dolphins and Montreal Gangs networks when edge restrictions are enforced. Recall that the proposed algorithms act in a greedy fashion by focusing on local decisions at each point. These results indicate that adding weight to any possible edge can result in adding weight to a locally good choice, but creates global difficulties in later iterations.

\subsection{Results for Edge Removal}
Here, we present the results of the edge removal for community preservation problem. The edge-anonymity prevented the integer programming formulation from making meaningful progress towards a solution, even with additional constraints enforcing that clusters are connected components. We present the number of edges remaining and run time on each network for all four rules in Table \ref{tbl:sparse_e}, with run time being reported in Table \ref{tbl:sparse_t}. For rule $2$, we run $100$ tests, and report the median, minimum and maximum of the number of edges remaining, and the mean and standard deviation of the run times. For the Les Mis data set, we only run $20$ tests due to lengthy solve time. Additionally, we expect that our method will remove all cross-cluster edges; we expect removing them will increase the modularity of an optimal partition. For all networks tested, we verified that removing such edges resulted in the resulting network having the same optimal partition as the initial network, allowing us the apply our method to these reduced networks. We test our method on these reduced networks, with number of edges remaining presented in Table \ref{tbl:sparse_e_p} and run times presented in Table \ref{tbl:sparse_t_p}. We report the number of edges remaining after removing cross-cluster edges as Pre, and the number of edges determined by Theorem \ref{thrm:star} in the column LB.
\begin{table}[h]
 \begin{center}
 \caption{Comparison of Number of Edges Remaining in Tested Networks}
 \resizebox{\textwidth}{!}{
  \begin{tabular}{|c|c|c|c|c|c|c|} 
  \hline
  Network & \#Edges & Rule 1 & Rule 2 Med (Min, Max) & Rule 3 & Rule 4 & LB\\ [0.5ex]
  \hline\hline
  Sawmill & 62 & 46 & 34 (33, 46) & 34 & 34 & 32\\
  \hline
  Dolphins small & 70 & 53 & 43 (38, 68) & 42 & 42 & 34\\
  \hline
  Karate & 78 & 31 & 33 (30, 36) & 31 & 31 & 30\\ 
  \hline
  Les Mis & 254 & 137 & 74 (72, 140) & 137 & 137 & 71 \\ 
  \hline\hline
  Ciel & 35 & 26 & 26 (25, 28) & 25 & 25 & 22\\
  \hline
  Caviar6 & 47 & 23 & 23 (23, 35) & 23 & 23 & 23\\
  \hline
  Rhodes & 66 & 21 & 23 (21, 27) & 21 & 21 & 19\\
  \hline
  Montreal Gangs & 75 & 35 & 32 (29, 36) & 35 & 36 & 26\\
  \hline
  Italian Gangs & 113 & 62 & 64 (62, 72) & 64 & 64 & 60\\ 
  \hline
  London Gangs & 133 & 44 & 45 (43, 88) & 44 & 44 & 42\\ 
  \hline
 \end{tabular}}
 \label{tbl:sparse_e}
 \end{center}
\end{table}

\begin{table}[h]
 \begin{center}
 \caption{Run Time (in seconds) on Tested Networks}
  \begin{tabular}{|c|c|c|c|c|} 
  \hline
  Network & Rule 1 & Rule 2 Mean (SD) & Rule 3 & Rule 4\\ [0.5ex] 
  \hline\hline
  Sawmill & 14.78 & 22.09 (8.40) & 23.96 & 23.76 \\
  \hline
  Dolphins small & 98.06 & 35.17 (8.69) & 24.67 & 23.89\\
  \hline
  Karate & 27.11 & 26.41 (13.37) & 23.06 & 23.09\\ 
  \hline
  Les Mis & 7200* & 5811.00 (1140.88) & 7200* & 7200*\\ 
  \hline\hline
  Ciel & 2.51 & 2.11 (0.46) & 1.70 & 1.69 \\
  \hline
  Caviar6 & 12.44 & 9.25 (2.90) & 11.54 & 11.59\\
  \hline
  Rhodes & 6.00 & 2.97 (0.84) & 2.72 & 2.64 \\
  \hline
  Montreal Gangs & 47.93 & 111.50 (57.73) & 172.28 & 177.58 \\
  \hline
  Italian Gangs & 837.89 & 1023.24 (235.80) & 1128.15 & 1156.27 \\ 
  \hline
  London Gangs & 334.29 & 1310.91 (1377.77) & 628.27 & 640.23 \\ 
  \hline
 \end{tabular}
 \label{tbl:sparse_t}
 \end{center}
\end{table}

\begin{table}[h]
 \begin{center}
 \caption{Comparison of Number of Edges Remaining in Tested Networks with Pre-Processing}
 \resizebox{\textwidth}{!}{
  \begin{tabular}{|c|c|c|c|c|c|c|c|} 
  \hline
  Network & \#Edges & Pre & Rule 1 & Rule 2 Med (Min, Max) & Rule 3 & Rule 4 & LB\\ [0.5ex] 
  \hline\hline
  Sawmill & 62 & 51 & 46 & 34 (33, 43) & 34 & 34 & 32\\
  \hline
  Dolphins small & 70 & 63 & 42 & 42 (38, 57) & 42 & 42 & 34\\
  \hline
  Karate & 78 & 57 & 31 & 33 (30, 40) & 30 & 30 & 30\\ 
  \hline
  Les Mis & 254 & 194 & 104 & 73 (72, 81) & 95 & 95 & 71 \\ 
  \hline\hline
  Ciel & 35 & 29 & 26 & 26 (25, 26) & 25 & 25 & 22\\
  \hline
  Caviar6 & 47 & 27 & 23 & 23 (23, 23) & 23 & 23 & 23\\
  \hline
  Rhodes & 66 & 46 & 23 & 23 (21, 28) & 21 & 21 & 19\\
  \hline
  Montreal Gangs & 75 & 44 & 34 & 33 (29, 37) & 33 & 33 & 26\\
  \hline
  Italian Gangs & 113 & 87 & 68 & 68 (62, 73) & 62 & 62 & 60\\ 
  \hline
  London Gangs & 133 & 90 & 59 & 44 (43, 49) & 44 & 44 & 42\\ 
  \hline
 \end{tabular}}
 \label{tbl:sparse_e_p}
 \end{center}
\end{table}

\begin{table}[h]
 \begin{center}
 \caption{Run Time (in seconds) on Tested Networks with Pre-Processing}
  \begin{tabular}{|c|c|c|c|c|} 
  \hline
  Network & Rule 1 & Rule 2 Mean (SD) & Rule 3 & Rule 4\\ [0.5ex] 
  \hline\hline
  Sawmill & 12.76 & 11.05 (1.39) & 10.78 & 10.75\\
  \hline
  Dolphins small & 18.04 & 20.63 (3.61) & 19.95 & 19.97 \\
  \hline
  Karate & 8.78 & 8.87 (0.93) & 10.03 & 9.86 \\ 
  \hline
  Les Mis & 1696.14 & 1253.12 (182.17) & 2086.95 & 1777.45 \\ 
  \hline\hline
  Ciel & 1.46 & 1.49 (0.13) & 1.42 & 1.45 \\
  \hline
  Caviar6 & 2.18 & 2.42 (0.37) & 2.50 & 2.31\\
  \hline
  Rhodes & 1.76 & 1.65 (0.30) & 1.61 & 1.51\\
  \hline
  Montreal Gangs & 11.75 & 12.25 (1.40) & 11.29 & 11.74\\
  \hline
  Italian Gangs & 308.31 & 349.39 (37.99) & 285.99 & 295.70 \\ 
  \hline
  London Gangs & 59.17 & 67.19 (9.26) & 68.42 & 71.46 \\ 
  \hline
 \end{tabular}
 \label{tbl:sparse_t_p}
 \end{center}
\end{table}

Here we see that for almost every data set, the rules ordering the edges based on contribution to modularity either perform as well as or better than the other rules. The minimums obtained by the second rule are at least as good as these rules, but the medians rarely outperform the other rules. Additionally, the high standard deviations in run time make it less appealing, even when it is the fastest rule. The second rule only consistently outperforms the third and fourth rules in the Les Mis data set. This is likely due to being able to identify orderings that remove problematic edges that prevent the removal of more edges in later iterations. We note that the first rule only happens to work well when the ordering of the edges is already satisfactory. While it performs well here, it is not guaranteed to do so in general, such as with the Dolphins data set, where the third and fourth rule will be consistent regardless of initial edge ordering. We note that, on these networks, the dynamic reordering impacts execution time negligibly, but does not improve the results. Additionally, by applying pre-processing to remove edges that cross clusters, we can remove at least $10\%$ of the edges in each network while still maintaining the same optimal partition. This reduction significantly improves the solve time in every network. Additionally, applying our algorithm to these reduced networks consistently improved the quality of the solutions found, and decreased the range of solutions identified by the second rule.

\section{Conclusions and Future Work}
We define two new problems regarding the modularity of a network. In the first problem, we seek to find the minimum set of edges (or weights) to add to a network to enforce that a given partition maximizes modularity. In the second problem, we seek to find the minimum set of edges in the network such that a partition optimizing modularity of the reduced network and the full network is the same. We provide an integer programming framework for these problems, and augment the method with the generation of disjunctive cuts and additional partitions. We demonstrate how the counter-intuitive behavior of modularity proves problematic for the branch-and-bound procedure and prevents proving nice properties about how adding edges impacts modularity. The integer programming formulation is able to solve the first problem when the ground truth communities are relatively distinct or similar to the initial optimal clusters, demonstrating that the difficulty of the problem is not just dependent on the size of the network. We also devise heuristics that are able to find high quality solutions to the problem significantly quicker than the integer programming formulation, and find solutions to the problem on instances that the integer programming formulation is not able to solve. For the second problem, our heuristics find that we are able to identify a small number of edges necessary to maintain the original community structure, often within a small percentage of edges away from identifying star networks within each cluster. 

Future work includes extending these methods to work for other optimization based clustering measures, such as modularity density, as well as the development of heuristics for these measures \cite{chen2013measuring}. Additional future work includes improving the components of the integer programming framework, such as symmetry breaking techniques in the branch-and-bound procedure. Improving the tightness of the McCormick inequalities in the weighted problem when enforcing a given partition is optimal as compared to previously visited partitions is also important to being able to apply the integer program to weighted networks. This framework can also be improved upon to utilize more advanced means of solving for a partition that optimizes modularity, allowing for these problems to be solved more efficiently and on larger networks.

\section*{Acknowledgements}
This work was supported by the U.S. Department of Homeland Security under Grant Award Number 2017-ST061-CINA01. The views and conclusions contained in this document are those of the authors and should not be interpreted as necessarily representing the official policies, either expressed or implied, of the U.S. Department of Homeland Security.

\bibliographystyle{abbrv-networks} 
\bibliography{ref.bib} 

\begin{thebibliography}{10}

\bibitem{mod_IP}
G.~Agarwal and D.~Kempe, {\em Modularity-maximizing graph communities via
  mathematical programming},  Eur. Physical J. B {\bf 66} (2008),  409--418.

\bibitem{aloise2010column}
D.~Aloise, S.~Cafieri, G.~Caporossi, P.~Hansen, S.~Perron, and L.~Liberti, {\em
  Column generation algorithms for exact modularity maximization in networks},
  Physical Review E {\bf 82} (2010),  046112.

\bibitem{anzoom2021review}
R.~Anzoom, R.~Nagi, and C.~Vogiatzis, {\em A review of research in illicit
  supply-chain networks and new directions to thwart them}, IISE Trans. (2021),
   1--59, https://doi.org/10.1080/24725854.2021.1939466.

\bibitem{Bahulkar2018}
A.~{Bahulkar}, B.K. {Szymanski}, N.O. {Baycik}, and T.C. {Sharkey}, {\em
  Community detection with edge augmentation in criminal networks}, 2018
  IEEE/ACM International Conference on Advances in Social Networks Analysis and
  Mining (ASONAM), 2018, pp.  1168--1175.

\bibitem{pajek}
V.~Batagelj and A.~Mrvar, {\em Pajek datasets},
  \url{http://vlado.fmf.uni-lj.si/pub/networks/data/esna/default.htm}, 2006.

\bibitem{berlusconi2013all}
G.~Berlusconi, {\em Do all the pieces matter? {A}ssessing the reliability of
  law enforcement data sources for the network analysis of wire taps}, Global
  Crime {\bf 14} (2013),  61--81.

\bibitem{berlusconi2016link}
G.~Berlusconi, F.~Calderoni, N.~Parolini, M.~Verani, and C.~Piccardi, {\em Link
  prediction in criminal networks: A tool for criminal intelligence analysis},
  PloS one {\bf 11} (2016),  e0154244.

\bibitem{blondel2008fast}
V.D. Blondel, J.L. Guillaume, R.~Lambiotte, and E.~Lefebvre, {\em Fast
  unfolding of communities in large networks}, J.  Stat. Mechanics: Theory
  Experiment {\bf 2008} (2008),  P10008.

\bibitem{brandes2007modularity}
U.~Brandes, D.~Delling, M.~Gaertler, R.~Gorke, M.~Hoefer, Z.~Nikoloski, and
  D.~Wagner, {\em On modularity clustering}, IEEE Trans.  Knowledge  Data Eng.
  {\bf 20} (2007),  172--188.

\bibitem{bright2013dismantling}
D.A. Bright, C.~Greenhill, and N.~Levenkova, ``{\em Dismantling criminal
  networks: Can node attributes play a role?}," {\em Crime and Networks},
  Routledge, \url{https://doi.org/10.4324/9781315885018}, 2013, pp.  160--174.

\bibitem{calderoni2017communities}
F.~Calderoni, D.~Brunetto, and C.~Piccardi, {\em Communities in criminal
  networks: A case study}, Social Networks {\bf 48} (2017),  116--125.

\bibitem{calderoni2020robust}
F.~Calderoni, S.~Catanese, P.~De~Meo, A.~Ficara, and G.~Fiumara, {\em Robust
  link prediction in criminal networks: A case study of the {S}icilian
  {M}afia}, Expert Syst. with Appl. {\bf 161} (2020),  113666.

\bibitem{casos}
CASOS, {\em Public datasets},
  \url{http://www.casos.cs.cmu.edu/tools/datasets/external/index.php}, 1986.

\bibitem{chen2014community}
M.~Chen, K.~Kuzmin, and B.K. Szymanski, {\em Community detection via
  maximization of modularity and its variants}, IEEE Trans.  Comput. Social
  Syst. {\bf 1} (2014),  46--65.

\bibitem{chen2013measuring}
M.~Chen, T.~Nguyen, and B.K. Szymanski, {\em On measuring the quality of a
  network community structure}, Proceedings of the 2013 International
  Conference on Social Computing, 2013, pp.  122--127.

\bibitem{cinar2017analyzing}
M.S. Cinar, B.~Genc, H.~Sever, and V.V. Raghavan, {\em Analyzing structure of
  terrorist networks by using graph metrics}, Proceedings of the 2017 IEEE
  International Conference on Big Knowledge (ICBK), 2017, pp.  9--16.

\bibitem{igraph}
G.~Csardi, T.~Nepusz, et~al., {\em The igraph software package for complex
  network research}, InterJournal, complex systems {\bf 1695} (2006),  1--9.

\bibitem{cuganesan2011developments}
S.~Cuganesan and D.~Lacey, {\em Developments in public sector performance
  measurement: {A} project on producing return on investment metrics for law
  enforcement}, Financial Accountability  Manage. {\bf 27} (2011),  458--479.

\bibitem{cunningham2016understanding}
D.~Cunningham, S.~Everton, and P.~Murphy, {\em Understanding dark networks: A
  strategic framework for the use of social network analysis}, Rowman \&
  Littlefield, Maryland, 2016.

\bibitem{deeb2019understanding}
J.~Deeb-Swihart, A.~Endert, and A.~Bruckman, {\em Understanding law enforcement
  strategies and needs for combating human trafficking}, Proceedings of the
  2019 CHI Conference on Human Factors in Computing Systems, 2019, pp.  1--14.

\bibitem{descormiers2011alliances}
K.~Descormiers and C.~Morselli, {\em Alliances, conflicts, and contradictions
  in {M}ontreal’s street gang landscape}, Int. Criminal Justice Review {\bf
  21} (2011),  297--314.

\bibitem{elsisy2020synthetic}
A.~Elsisy, A.~Mandviwalla, B.~Szymanski, and T.~Sharkey, {\em A synthetic
  network generator for covert network analytics}, arXiv preprint
  arXiv:2008.04445 (2020).

\bibitem{monotone}
M.~Fischetti, I.~Ljubic, M.~Monaci, and M.~Sinnl, {\em Interdiction games and
  monotonicity, with application to knapsack problems}, INFORMS J.  Comput.
  {\bf 31} (2019),  390--410.

\bibitem{gill2013dynamic}
J.~Gill and J.R. Freeman, {\em Dynamic elicited priors for updating covert
  networks}, Network Sci. {\bf 1} (2013),  68--94.

\bibitem{goldstein2020exploitative}
R.~Goldstein, M.W. Sances, and H.Y. You, {\em Exploitative revenues, law
  enforcement, and the quality of government service}, Urban Affairs Review
  {\bf 56} (2020),  5--31.

\bibitem{grund2015ethnic}
T.U. Grund and J.A. Densley, {\em Ethnic homophily and triad closure: Mapping
  internal gang structure using exponential random graph models}, J.
  Contemporary Criminal Justice {\bf 31} (2015),  354--370.

\bibitem{holmes2008minority}
M.D. Holmes, B.W. Smith, A.B. Freng, and E.A. Mu{\~n}oz, {\em Minority threat,
  crime control, and police resource allocation in the {S}outhwestern {U}nited
  {S}tates}, Crime  Delinquency {\bf 54} (2008),  128--152.

\bibitem{studio2017cplex}
{IBM ILOG CPLEX Optimization Studio}, {\em User's manual for {CPLEX}, version
  20.1},
  \url{https://www.ibm.com/docs/en/icos/20.1.0?topic=cplex-users-manual}, 2020.

\bibitem{kennedy2011risk}
L.W. Kennedy, J.M. Caplan, and E.~Piza, {\em Risk clusters, hotspots, and
  spatial intelligence: {R}isk terrain modeling as an algorithm for police
  resource allocation strategies}, J.  Quantitative Criminology {\bf 27}
  (2011),  339--362.

\bibitem{knuth1993stanford}
D.E. Knuth, {\em The Stanford GraphBase: {A} platform for combinatorial
  computing}, ACM Press, New York, 1993.

\bibitem{lusseau2003bottlenose}
D.~Lusseau, K.~Schneider, O.J. Boisseau, P.~Haase, E.~Slooten, and S.M. Dawson,
  {\em The bottlenose dolphin community of {D}oubtful {S}ound features a large
  proportion of long-lasting associations}, Behavioral Ecology  Sociobiology
  {\bf 54} (2003),  396--405.

\bibitem{malm2011networks}
A.~Malm and G.~Bichler, {\em Networks of collaborating criminals: Assessing the
  structural vulnerability of drug markets}, J.  Res.  Crime  Delinquency {\bf
  48} (2011),  271--297.

\bibitem{linkpred}
V.~Mart{\'\i}nez, F.~Berzal, and J.C. Cubero, {\em A survey of link prediction
  in complex networks}, ACM Comput. Surveys (CSUR) {\bf 49} (2016),  1--33.

\bibitem{mccormick1976computability}
G.P. McCormick, {\em Computability of global solutions to factorable nonconvex
  programs: {P}art {I}—{C}onvex underestimating problems}, Math. Program.
  {\bf 10} (1976),  147--175.

\bibitem{michael1997modeling}
J.H. Michael and J.G. Massey, {\em Modeling the communication network in a
  sawmill}, Forest Products J. {\bf 47} (1997),  25--30.

\bibitem{mohanty2020computational}
P.~Mohanty, {\em A computational approach to identify covertness and collusion
  in social networks}, Ph.D.~thesis, University of Minnesota, 2020.

\bibitem{caviar6}
C.~Morselli, {\em Inside criminal networks}, Springer, 2009.

\bibitem{natarajan2000understanding}
M.~Natarajan, {\em Understanding the structure of a drug trafficking
  organization: {A} conversational analysis}, Crime Prevention Studies {\bf 11}
  (2000),  273--298.

\bibitem{natarajan2006understanding}
M.~Natarajan, {\em Understanding the structure of a large heroin distribution
  network: A quantitative analysis of qualitative data}, J.  Quantitative
  Criminology {\bf 22} (2006),  171--192.

\bibitem{newman2003random}
M.E. Newman, {\em Random graphs as models of networks}, Handbook  Graphs
  Networks {\bf 1} (2003),  35--68.

\bibitem{newman2004fast}
M.E. Newman, {\em Fast algorithm for detecting community structure in
  networks}, Physical Review E {\bf 69} (2004),  66133.

\bibitem{newman2016equivalence}
M.E. Newman, {\em Equivalence between modularity optimization and maximum
  likelihood methods for community detection}, Physical Review E {\bf 94}
  (2016),  52315.

\bibitem{mod_orig}
M.E. Newman and M.~Girvan, {\em Finding and evaluating community structure in
  networks}, Physical Review E {\bf 69} (2004),  26113.

\bibitem{ostrowski2011orbital}
J.~Ostrowski, J.~Linderoth, F.~Rossi, and S.~Smriglio, {\em Orbital branching},
  Math. Program. {\bf 126} (2011),  147--178.

\bibitem{owen2002value}
J.H. Owen and S.~Mehrotra, {\em On the value of binary expansions for general
  mixed-integer linear programs}, Oper. Res. {\bf 50} (2002),  810--819.

\bibitem{pemmaraju2003computational}
S.~Pemmaraju and S.~Skiena, {\em Computational discrete mathematics:
  Combinatorics and graph theory with {M}athematica{\textregistered}},
  Cambridge University Press, New York, 2003.

\bibitem{rhodes2009inferring}
C.~Rhodes and P.~Jones, {\em Inferring missing links in partially observed
  social networks}, J.   Oper. Res. Soc. {\bf 60} (2009),  1373--1383.

\bibitem{rhodes2007social}
C.J. Rhodes and E.~Keefe, {\em Social network topology: {A} {B}ayesian
  approach}, J.   Oper. Res. Soc. {\bf 58} (2007),  1605--1611.

\bibitem{serrano2021community}
B.~Serrano and T.~Vidal, {\em Community detection in the stochastic block model
  by mixed integer programming}, arXiv preprint arXiv:2101.12336 (2021).

\bibitem{spapens2011interaction}
T.~Spapens, {\em Interaction between criminal groups and law enforcement: {T}he
  case of ecstasy in the {N}etherlands}, Global crime {\bf 12} (2011),  19--40.

\bibitem{sparrow1991application}
M.K. Sparrow, {\em The application of network analysis to criminal
  intelligence: An assessment of the prospects}, Social Networks {\bf 13}
  (1991),  251--274.

\bibitem{strang2014network}
S.J. Strang, ``{\em Network analysis in criminal intelligence}," {\em Networks
  and network analysis for defence and security}, Springer,
  \url{https://doi.org/10.1007/978-3-319-04147-6_1}, 2014, pp.  1--26.

\bibitem{modSurvey}
K.~Taha, {\em Static and dynamic community detection methods that optimize a
  specific objective function: A survey and experimental evaluation}, IEEE
  Access {\bf 8} (2020),  98330--98358.

\bibitem{van2009introduction}
R.C. Van~der Hulst, {\em Introduction to {S}ocial {N}etwork {A}nalysis ({SNA})
  as an investigative tool}, Trends  Organized Crime {\bf 12} (2009),
  101--121.

\bibitem{zachary}
W.W. Zachary, {\em An information flow model for conflict and fission in small
  groups}, J.  Anthropological Res. {\bf 33} (1977),  452--473.

\end{thebibliography}

\newpage
\begin{appendices}
\section{Appendix A: Computation of Change in Modularity}
\label{app:proofs}
\subsection{Change in Modularity from Including an Edge Within a Cluster}
\begin{proof}
Consider adding edge $e = (u,v)$ where $u, v \in C_1$. Let $z^e$ be the vector such that $z_e^e = 1$ and $z_{e'}^e = 0$ for $e' \ne e$. We compute the difference $(Q_T(z^e) - Q_T(0)$. To simplify fractions, we multiply by $(2m)^2 (2m+2)^2$.

\begin{align*}
    & (2m)^2(2m+2)^2(Q_T (z^e) - Q_T (0)) \\
    & = (2m)^2(2m+2) A_{uu} - (2m)^2 (d_u+1)^2 - (2m)(2m+2)^2 A_{uu} + (2m+2)^2 d_u^2 \\
    & + (2m)^2(2m+2) A_{vv} - (2m)^2 (d_v+1)^2 - (2m)(2m+2)^2 A_{vv} + (2m+2)^2 d_v^2 \\
    & + 2\left((2m)^2(2m+2) A_{uv} + (2m)^2(2m+2) - (2m)^2(d_u+1)(d_v+1)\right) \\& + \left(- (2m)(2m+2)^2 A_{uv} + (2m+2)^2 d_u d_v \right)\\
    & + 2\left(\sum_{u,v \ne x \in C_1} \left((2m)^2(2m+2)A_{ux} - (2m)^2(d_u+1)d_x - (2m)(2m+2)^2 A_{ux} + (2m+2)^2 d_u d_x \right)\right) \\
    & + 2\left(\sum_{u,v \ne y \in C_1} \left((2m)^2(2m+2)A_{vy} - (2m)^2(d_v+1)d_y - (2m)(2m+2)^2 A_{vy} + (2m+2)^2 d_v d_y\right) \right) \\
    & + \sum_{u, v \ne x,y\in C_1} \left((2m)^2(2m+2)A_{xy} - (2m)^2 d_x d_y - (2m)(2m+2)^2 A_{xy} + (2m+2)^2 d_x d_y \right) \\
    & + \sum_{C_i \in T, i \ne 1} \sum_{x,y \in C_i} \left((2m)^2(2m+2)A_{xy} - (2m)^2 d_x d_y - (2m)(2m+2)^2 A_{xy} + (2m+2)^2 d_x d_y \right) \\
    & = (-8m^2 - 8m)A_{uu} - 8m^2 d_u - 4m^2 +8m d^2_u + 4d^2_u\\
    & + (-8m^2 - 8m)A_{vv} - 8m^2 d_v - 4m^2 +8m d^2_v + 4d^2_v\\
    & + 2\left((-8m^2 - 8m)A_{uv} + 8m^3 + 8m^2 - (4m^2)(d_u d_v + d_u + d_v +1) + (4m^2 + 8m + 4) d_u d_v \right)\\
    & + 2\left(\sum_{u,v \ne x \in C_1} \left( (-8m^2 -8m)A_{ux} - 4m^2(d_u d_x + d_x) + (4m^2 + 8m +4) d_u d_x  \right)\right)\\
    & + 2\left(\sum_{u,v \ne y \in C_1} \left( (-8m^2 -8m)A_{vy} - 4m^2(d_v d_y + d_y) + (4m^2 + 8m +4) d_v d_y  \right)\right)\\
    & + \sum_{u, v \ne x,y\in C_1} \left((-8m^2 -8m)A_{xy} + (8m +4) d_x d_y \right) \\& + \sum_{C_i \in T, i \ne 1} \sum_{x,y\in C_i} \left((-8m^2 -8m)A_{xy} + (8m +4) d_x d_y \right)\\
    & = (-8m^2 - 8m)A_{uu} + (8m +4) d^2_u - 8m^2 d_u - 4m^2\\
    & + (-8m^2 - 8m)A_{vv} + (8m +4) d^2_v - 8m^2 d_v - 4m^2\\
    & + 2\left((-8m^2 - 8m)A_{uv}  + (8m +4) d_u d_v + 8m^3 + 4m^2 - 4m^2 d_u -4m^2 d_v \right)\\
    & + 2\left(\sum_{u,v \ne x \in C_1} \left( (-8m^2 -8m)A_{ux} + (8m +4) d_u d_x - 4m^2 d_x \right)\right)\\
    & + 2\left(\sum_{u,v \ne y \in C_1} \left( (-8m^2 -8m)A_{vy} + (8m+ 4)d_v d_y - 4m^2 dy  \right)\right)\\
    & + \sum_{u, v \ne x,y\in C_1} \left((-8m^2 -8m)A_{xy} + (8m +4) d_x d_y \right)\\& + \sum_{C_i \in T, i \ne 1} \sum_{x,y\in C_i} \left((-8m^2 -8m)A_{xy} + (8m +4) d_x d_y \right) \\
    &\\
    & = 16m^3 - \sum_{x \in C_1} 16m^2 d_x + \sum_{C_i \in T} \sum_{x,y\in C_i} \left((-8m^2 -8m)A_{xy} + (8m +4) d_x d_y \right)
\end{align*}

Thus, the increase in modularity by adding $(u,v)$ is $$\frac{16m^3 - \sum_{x \in C_1} 16m^2 d_x + \sum_{C_i \in T} \sum_{x,y\in C_i} \left((-8m^2 -8m)A_{xy} + (8m +4) d_x d_y \right)}{(2m)^2 (2m+2)^2}\text{.}$$
\end{proof}

\newpage
\subsection{Change in Modularity from Including an Edge Between Clusters}
\begin{proof}
Consider adding edge $e = (u,v)$ where $u\in C_1$ and $v \in C_2$. Let $z^e$ be the vector such that $z_e^e = 1$ and $z_{e'}^e = 0$ for $e' \ne e$.  We compute the difference $(Q_T(z^e) - Q_T(0)$. We simplify fractions by multiplying by $(2m)^2 (2m+2)^2$.

\begin{align*}
    & (2m)^2(2m+2)^2(Q_T (z^e) - Q_T (0)) \\
    & = (2m)^2(2m+2) A_{uu} - (2m)^2 (d_u+1)^2 - (2m)(2m+2)^2 A_{uu} + (2m+2)^2 d_u^2 \\
    & + 2\left(\sum_{u \ne x \in C_1} \left((2m)^2(2m+2)A_{ux} - (2m)^2(d_u+1)d_x - (2m)(2m+2)^2 A_{ux} + (2m+2)^2 d_u d_x \right)\right) \\
    & + \sum_{u \ne x,y\in C_1} \left((2m)^2(2m+2)A_{xy} - (2m)^2 d_x d_y - (2m)(2m+2)^2 A_{xy} + (2m+2)^2 d_x d_y \right) \\
    & + (2m)^2(2m+2) A_{vv} - (2m)^2 (d_v+1)^2 - (2m)(2m+2)^2 A_{vv} + (2m+2)^2 d_v^2 \\
    & + 2\left(\sum_{v \ne y \in C_2} \left((2m)^2(2m+2)A_{vy} - (2m)^2(d_v+1)d_y - (2m)(2m+2)^2 A_{vy} + (2m+2)^2 d_v d_y\right) \right) \\
    & + \sum_{v \ne x,y\in C_2} \left((2m)^2(2m+2)A_{xy} - (2m)^2 d_x d_y - (2m)(2m+2)^2 A_{xy} + (2m+2)^2 d_x d_y \right) \\
    & + \sum_{C_i \in T, i \notin \{1,2 \}} \sum_{x,y \in C_i} \left((2m)^2(2m+2)A_{xy} - (2m)^2 d_x d_y - (2m)(2m+2)^2 A_{xy} + (2m+2)^2 d_x d_y \right)
    &\\
    & = (-8m^2 - 8m)A_{uu} - 8m^2 d_u - 4m^2 +8m d^2_u + 4d^2_u\\
    & + 2\left(\sum_{u \ne x \in C_1} \left( (-8m^2 -8m)A_{ux} - 4m^2(d_u d_x + d_x) + (4m^2 + 8m +4) d_u d_x  \right)\right)\\
    &+ \sum_{u \ne x,y\in C_1} \left((-8m^2 -8m)A_{xy} + (8m +4) d_x d_y \right) \\
    & + (-8m^2 - 8m)A_{vv} - 8m^2 d_v - 4m^2 +8m d^2_v + 4d^2_v\\
    & + 2\left(\sum_{v \ne y \in C_2} \left( (-8m^2 -8m)A_{vy} - 4m^2(d_v d_y + d_y) + (4m^2 + 8m +4) d_v d_y  \right)\right)\\
    & + \sum_{v \ne x,y\in C_2} \left((-8m^2 -8m)A_{xy} + (8m +4) d_x d_y \right) \\& + \sum_{C_i \in T, i \notin \{1,2 \}} \sum_{ x,y\in C_i} \left((-8m^2 -8m)A_{xy} + (8m +4) d_x d_y \right)\\
    & = (-8m^2 - 8m)A_{uu} + (8m +4) d^2_u - 8m^2 d_u - 4m^2\\
    & + 2\left(\sum_{u \ne x \in C_1} \left( (-8m^2 -8m)A_{ux} + (8m +4) d_u d_x - 4m^2 d_x \right)\right)\\
    & + \sum_{u \ne x,y\in C_1} \left((-8m^2 -8m)A_{xy} + (8m +4) d_x d_y \right)\\
    & + (-8m^2 - 8m)A_{vv} + (8m +4) d^2_v - 8m^2 d_v - 4m^2\\
    & + 2\left(\sum_{v \ne y \in C_2} \left( (-8m^2 -8m)A_{vy} + (8m+ 4)d_v d_y - 4m^2 dy  \right)\right)\\
    & + \sum_{v \ne x,y\in C_2} \left((-8m^2 -8m)A_{xy} + (8m +4) d_x d_y \right) \\& + \sum_{C_i \in T, i \notin \{1,2 \}} \sum_{x,y\in C_i} \left((-8m^2 -8m)A_{xy} + (8m +4) d_x d_y \right) \\
    &\\
    & = -8m^2 - \sum_{x \in C_1} 8m^2 d_x - \sum_{y \in C_2} 8m^2 d_y +\sum_{C_i \in T} \sum_{x,y\in C_i} \left((-8m^2 -8m)A_{xy} + (8m +4) d_x d_y \right)
\end{align*}

Thus, the increase in modularity by adding $(u,v)$ is $$\frac{-8m^2 - \sum_{x \in C_1} 8m^2 d_x - \sum_{y \in C_2} 8m^2 d_y +\sum_{C_i \in T} \sum_{x,y\in C_i} \left((-8m^2 -8m)A_{xy} + (8m +4) d_x d_y \right)}{(2m)^2 (2m+2)^2}\text{.}$$
\end{proof}

\newpage
\section{Comparison of Master and Subproblem Solve Times}
\label{app:time}

\begin{table}[h!]
\begin{center}
\caption{Comparison of Run Times (in seconds) with Default CPLEX Symmetry Breaking}
\begin{tabular}{|c|c|c|c|c|}
\hline
Network        & Master (IP)        & Sub (IP) & Master (IP+)       & Sub (IP+) \\ \hline\hline
Sawmill        & 288.89             & 10.09    & 3050.45            & 5.78      \\ \hline
Dolphins       & \textgreater{}7200 & 27.81    & \textgreater{}7200 & 25.24     \\ \hline
Karate         & 514.06             & 12.56    & 51.39              & 8.30      \\ \hline
Les Mis        & 83.01              & 386.96   & 20.41              & 166.31    \\ \hline\hline
Ciel           & 1.13               & 0.20     & 1.09               & 0.19      \\ \hline
Caviar6        & 115.45             & 7.44     & 27.92              & 4.92      \\ \hline
Rhodes         & \textgreater{}7200 & 5.88     & \textgreater{}7200 & 4.20      \\ \hline
Montreal Gangs & \textgreater{}7200 & 24.64    & \textgreater{}7200 & 103.38    \\ \hline
Italian Gangs  & 5.06               & 17.02    & 4.53               & 13.33     \\ \hline
London Gangs   & 24.13              & 31.36    & 12.42              & 16.88     \\ \hline
\end{tabular}
\end{center}
\end{table}

\begin{table}[h!]
\begin{center}
\caption{Comparison of Run Times (in seconds) with Maximum CPLEX Symmetry Breaking}
\begin{tabular}{|c|c|c|c|c|}
\hline\hline
Network        & Master (IP)        & Sub (IP) & Master (IP+)       & Sub (IP+) \\ \hline
Sawmill        & 261.12             & 9.98     & 3265.61            & 5.75      \\ \hline
Dolphins       & \textgreater{}7200 & 15.64    & \textgreater{}7200 & 67.66     \\ \hline
Karate         & 432.62             & 12.52    & 51.31              & 7.94      \\ \hline
Les Mis        & 83.41              & 366.24   & 30.84              & 168.44    \\ \hline\hline
Ciel           & 1.11               & 0.20     & 1.13               & 0.20      \\ \hline
Caviar6        & 112.99             & 7.59     & 27.22              & 4.95      \\ \hline
Rhodes         & \textgreater{}7200 & 5.53     & \textgreater{}7200 & 1.94      \\ \hline
Montreal Gangs & \textgreater{}7200 & 25.14    & \textgreater{}7200 & 75.36     \\ \hline
Italian Gangs  & 4.84               & 17.36    & 4.81               & 14.75     \\ \hline
London Gangs   & 24.11              & 30.55    & 14.09              & 18.64     \\ \hline
\end{tabular}
\end{center}
\end{table}

\end{appendices}
\end{document}